\documentclass[aps, 10pt, prb, twocolumn,floatfix, superscriptaddress]{revtex4-2}

\usepackage{graphicx}
\usepackage{dcolumn}
\usepackage{bm}
\usepackage[mathlines]{lineno}

\usepackage{tikz}
\usetikzlibrary{decorations.pathmorphing}
\usepackage{graphicx} 
\usepackage{physics}
\usepackage[utf8]{inputenc}
\usepackage[T1]{fontenc}
\usepackage{float}
\usepackage[english]{babel}
\usepackage{graphicx}
\usepackage{epsfig}
\usepackage{epstopdf}
\usepackage{amsmath}
\usepackage{amssymb}
\usepackage{times}
\usepackage{setspace}
\usepackage{verbatim}
\usepackage{xcolor}
\usepackage{mathrsfs}
\usepackage{euscript}
\usepackage{mathptmx}
\usepackage{mathtools}
\usepackage{multirow}
\usepackage[normalem]{ulem}
\usepackage[colorlinks=true, linkcolor=blue,citecolor=red, urlcolor=blue, hypertexnames=true]{hyperref}
\newcommand{\bea}{\begin{eqnarray}}
\newcommand{\eea}{\end{eqnarray}}

\newcommand{\ii}{\text{i}}
\newcommand{\up}{\uparrow}
\newcommand{\dn}{\downarrow}
\newcommand{\U}{\mathcal{U}}
\newcommand{\I}{\mathbb{I}}
\newcommand{\svn}{S_{vN}}

\begin{document}

\title{Page curve like dynamics in Interacting Quantum Systems}

\date{\today}

\author{Tamoghna Ray}
\email{tamoghna.ray@icts.res.in}
\affiliation{International Centre for Theoretical Sciences, Tata Institute of Fundamental Research,
Bangalore 560089, India}

\author{Abhishek Dhar}
\email{abhishek.dhar@icts.res.in }
\affiliation{International Centre for Theoretical Sciences, Tata Institute of Fundamental Research,
Bangalore 560089, India}

\author{Manas Kulkarni}
\email{manas.kulkarni@icts.res.in}
\affiliation{International Centre for Theoretical Sciences, Tata Institute of Fundamental Research,
Bangalore 560089, India}

\begin{abstract}
We study the dynamics of entanglement in a one-dimensional $XXZ$ spin-$1/2$ chain, with and without integrability-breaking interactions, that is connected to a bath. We start from a state where the system and bath are completely unentangled, and the bath is polarized spin-down. We consider two different initial states for the system - (i) a polarized spin-up state, and (ii) an infinite temperature state. In the particle representation of the spin chain, the polarized spin-up state corresponds to a filled state, while the polarized spin-down state corresponds to an empty state. Starting from these inhomogeneous quenches, in all the above-mentioned cases we obtain the Page curve like behavior in the entanglement. We report different power-law behavior in the growth of entanglement for different initial states and different kinds of baths (interacting and non-interacting). In an attempt to explore plausible deep connections between entanglement and Boltzmann entropy, we investigate the latter in both the filled and the infinite temperature case, for the system and the bath. For the filled case, the Boltzmann entropy of the system has the form of a Page curve but quantitatively deviates from the entanglement. On the other hand, the entropy of the bath keeps increasing. Remarkably, for the infinite temperature case, we find that the system and bath Boltzmann entropies agree with the entanglement entropy, after and before the Page time, respectively. Our findings are expected to hold for generic interacting quantum systems and could be of relevance to black hole physics.
\end{abstract}

\maketitle


\section{Introduction}
The study of entanglement is one of the foundational problems in quantum mechanics \cite{JLDec2009,JEOct2006,TNSep2018}. The dynamics of sub-system entanglement, starting from a quench, in a quantum many-body system is an interesting problem. In a series of seminal works, Cardy and Calabrese laid the foundation for such investigations using conformal field theory \cite{PCJune2004}, where they show linear \cite{PCApr2005} and logarithmic \cite{PCOct2007} growth in the entanglement for non-interacting  systems. Several studies have been done in both non-interacting \cite{PCApr2005,PCOct2007,TADec2008,MF2010,VE2010,FS2011,VE2012,DJ2017,BB2018,XC2019,OG2020,OA2021,SS2021,FA2022,LC2023} and interacting systems \cite{TADec2008,VAAug2014,GMNov2017,VA2017,VE2017,MZ2020,VANov2021} which show that different kinds of quenches lead to different behavior. Broadly, two different kinds of quenches have been investigated, homogeneous \cite{DC2006,VA2018,XC2019} and inhomogeneous \cite{PCOct2007,VAAug2014,ML2017,BB2018}.

Such studies are also of interest in the context of black hole evaporation, leading to the famous black hole information paradox \cite{SM2009,AAJul2021,SR2021}. Hawking showed that black holes emit thermal radiation, also known as the Hawking radiation. His semiclassical calculations showed that the black hole, along with the vacuum outside, which is initially in a pure state, evolves into a thermal state, and its entanglement with the surrounding radiation field keeps increasing \cite{SH1975}. However, this seems to break  unitarity in quantum mechanics. It was argued by Page \cite{DPDec1993,DPSep2013}, that the entanglement between the black hole and the surrounding radiation field increases, due to radiation evaporating from the black hole. Page argued that this growth in entanglement occurs till the black hole has evaporated to half of its initial mass, at which point it is maximally entangled with the surrounding field. This time is known as the Page time. Any evaporation after this leads to a decrease in entanglement, and finally, when the entire black hole has evaporated into the surrounding radiation field, it is completely unentangled. This behavior of the entanglement is known as the Page curve. Several studies have realized this behavior in microscopic non-interacting models \cite{SKJune2024,JGApr2024,MSFeb2024,JGJan2025,KGJan2025}, where a system (modeling the black hole) is connected to a bath (modeling the surrounding radiation field). The evolution in such studies typically starts from a local (inhomogeneous) quench, and several interesting aspects such as the hydrodynamic behavior~\cite{MSFeb2024}, effects of local dephasing emulating either imperfections or interactions \cite{KGJan2025}, and Linbladian baths \cite{JGApr2024,JGJan2025} have been explored. In Refs.~\onlinecite{DN2021} and \onlinecite{KS2021}, the Page curve has been obtained in interacting SYK models. The role of interactions has recently been explored further \cite{RJFeb2025,HLFeb2025}. The effect of interactions is still not well understood, and it is an interesting problem to study the dynamics of entanglement in interacting systems. Additionally, integrability \cite{VAAug2014,VA2017,ML2017,VA2018} and the lack thereof \cite{HK2013,TRJune2019} play an important role in the dynamics of the particles in the system, which in turn affects the entanglement of the system. It was observed in Ref.~\onlinecite{MSFeb2024} that there is a remarkable agreement between the dynamics of the fine-grained entanglement entropy and a coarse-grained hydrodynamic entropy. The main goals of the present work are to understand the details of the entanglement entropy dynamics in an  
interacting system and also to examine if the agreement between the fine-grained entanglement entropy and the coarse-grained Boltzmann entropy continues to hold. 

In this work, we study the dynamics of entanglement in a one-dimensional spin-$1/2$ $XXZ$ chain, with and without integrability breaking interactions, that is connected to a bath that can itself be either an interacting integrable bath or a non-interacting bath. Using the standard Jordan-Wigner transformation, the model can be transformed into a particle model, and we will often use the notation of the particle representation. We start from an initial state where the system and bath are completely unentangled and the bath is completely empty. Such a setup corresponds to the one described in the Page setup, where the black hole is in a pure state and the surrounding field (rest of the universe), which acts as the bath, is completely empty. We consider two different initial states for the system - (i) a completely filled state, and (ii) an infinite temperature state. The first case corresponds to the same initial state as in Refs.~\onlinecite{SKJune2024,MSFeb2024,HLFeb2025}. The second initial state is more relevant to the black hole model considered by Page where the black hole is initially in a high thermodynamic (Boltzmann) entropy state. Starting from these inhomogeneous quenched states, in all the above mentioned cases we obtain the Page curve like behavior in the entanglement. We report different exponents in the growth of entanglement for different initial states and different kinds of baths. 
Secondly, we compute the Boltzmann entropy of the bath and the system. For this we divide the full setup into spatial cells and consider a coarse-grained description (a macrostate) specified by the average number of particles and the average energy in each cell. The Boltzmann entropy counts the number of microstates for this macrostate~\cite{de2006,tasaki2016, goldstein2017,mori2018,Pandey23}. Starting from the filled state, the Boltzmann entropy of the system shows a Page curve like behavior, while the Boltzmann entropy of the bath keeps increasing.
Remarkably, starting from an infinite temperature state, we find that the dynamics of the Boltzmann entropy of the system agrees with that of the entanglement entropy \emph{after} the Page time. On the other hand, the Boltzmann entropy of the bath agrees with the entanglement entropy \emph{before} the Page time. Furthermore, our findings indicate that the Boltzmann entropy of the bath increases indefinitely for an infinite bath.

This article is structured as follows. In Sec.~\ref{sec:model}, we describe the models and the initial states used in the study. In Sec.~\ref{sec:entropy}, we define the primary quantities of interest - the von Neumann entropy and the Boltzmann entropy. In Sec.~\ref{sec:early_time_analytics} we present some analytic results using perturbation theory for the early time behaviour of various observables. In Sec.~\ref{sec:numerical_results}, we discuss the numerical results based on matrix product state computations for integrable and non-integrable systems connected to an $XXZ$ bath. We conclude with a discussion in Sec.~\ref{sec:conclusions}. Numerical results for the case when the integrable system is connected to a non-interacting bath are reported in Appendix~\ref{app:non_int_bath}. Some results on the spectrum of the system-bath Hamiltonian and their role in explaining the freezing of the dynamics, observed in certain cases,  are discussed in Appendix~\ref {app:overlap}. In Appendix~\ref{app:TEBD} we discuss the details of the numerical algorithm used.

\section{Model}
\label{sec:model}
In this section, we describe the models and the initial states used in the study. For the system, we consider a one-dimensional spin-$1/2$ $XXZ$ chain with $L_S$ sites. The Hamiltonian of the system is given by
\begin{align}
    H_{\rm sys} &= J\sum_{i= -(L_S-1)}^{-1}\left(S^x_{i}S^x_{i+1} + S^y_{i}S^y_{i+1} + \Delta S^z_{i}S^z_{i+1}\right) \nonumber\\
    &+ J'\sum_{i=-(L_S-1)}^{-2} S^z_{i}S^z_{i+2},
    \label{eq:system_hamiltonian}
\end{align}
where $J$ and $J'$ are the nearest-neighbor (NN) and next nearest-neighbor (NNN) coupling strengths, and $\Delta$ is the $z$-anisotropy. In the Fermionic language, the $\Delta$ maps to the interaction strength. $S^{x,y,z}_i = \sigma^{x,y,z}_i/2$ where $\sigma^{x,y,z}$ are the Pauli matrices at the $i^{th}$ site. For $J' = 0$, the system is integrable. The system is initialized in two different kinds of states. In the first case, the system is filled (polarized spin-up state),
\begin{equation}
    \ket{\psi_{\rm sys}(0)} = \ket{\up,\dots,\up}.
    \label{eq:sys_pol_init}
\end{equation}
In the second case, the system is initialized in a high Boltzmann entropy state (see Appendix~\ref{app:TEBD}) . 

\begin{figure}
    \centering
    \includegraphics[width=1.0\linewidth]{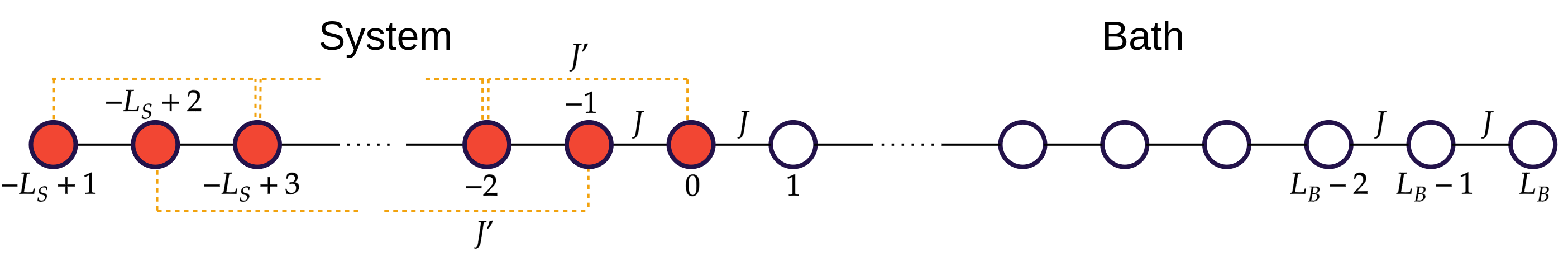}
    \caption{Schematic diagram of the model. The system, denoted by red circles, is a one-dimensional spin-$1/2$ $XXZ$ chain with $L_S$ sites. The system is connected to a bath of size $L_B$ $(L_B>> L_S)$, denoted by empty circles, on one end. The nearest neighbor couplings are denoted by solid lines and the next nearest couplings are denoted by dashed lines. The system, bath, and the system-bath Hamiltonians are given by Eqs.~\eqref{eq:system_hamiltonian}, \eqref{eq:int_bath_hamil}, and \eqref{eq:int_sys_bath} respectively.}
    \label{fig:schematic}
\end{figure}

The system is connected to a bath at one end (see Fig.~\ref{fig:schematic} for a schematic diagram of the setup).
The bath is taken to be an interacting XXZ spin chain with $L_B$ sites with $L_B\gg L_S$. 
The bath and system-bath Hamiltonians are given by
\begin{align}
    \label{eq:int_bath_hamil}
    H_{\rm bath} &= J\sum_{i=1}^{L_B-1}\left(S^x_{i}S^x_{i+1} + S^y_{i}S^y_{i+1} + \Delta S^z_{i}S^z_{i+1}\right),\\
     H_{\rm sys-bath} &= J\left(S^x_{0}S^x_{1} + S^y_{0}S^y_{1} + \Delta S^z_{0}S^z_{1}\right).     
    \label{eq:int_sys_bath}
\end{align}
We also discuss the case where the bath is taken to be non-interacting in Appendix~\ref {app:non_int_bath}, which corresponds to setting $\Delta = 0$ in Eqs.~\eqref{eq:int_bath_hamil} and \eqref{eq:int_sys_bath}. The Hamiltonian of the full chain is given by
\begin{equation}
    H = H_{\rm sys} + H_{\rm sys-bath} + H_{\rm bath}.
    \label{eq:sys_micro_bath_hamil}
\end{equation}
For the bath, we always start from an initial condition corresponding to an empty state (polarized spin-down state),
\begin{equation}
    \ket{\psi_{\rm bath}(0)} = \ket{\dn,\dots,\dn}.
    \label{eq:micro_bath_pol_init}
\end{equation}

\section{Entanglement and Boltzmann entropy}
\label{sec:entropy}
In this section, we define the primary quantities of interest. We investigate the behavior of two kinds of entropies - (i) von Neumann entropy, and (ii) Boltzmann entropy. The von Neumann entanglement entropy between two bipartitions of a system in a pure state $\rho = \ket{\psi}\bra{\psi}$ is given by
\begin{eqnarray}
    S_{vN} &=& -\Tr_A\rho_A \ln \rho_A =  -\Tr_B \rho_B \ln \rho_B, 
    \label{eq:SvN}
\end{eqnarray}
where $A$ and $B$ are the two bipartitions of the system and $\rho_{A/B} = \Tr_{B/A} \rho$. For our study, the two bipartitions are given by the system and the bath,  and the entanglement entropy computed is the system-bath entanglement.

The Boltzmann entropy is a coarse-grained entropy~\cite{de2006,tasaki2016, goldstein2017,mori2018,Pandey23}, and for this, we first need to define our coarse-grained description (macrostate) of the full setup. 
The macrostate we consider is one where we divide the full setup into spatial cells of size $L_S$ and specify the average number of particles (or magnetization) and the average energy in each cell. The Boltzmann entropy then essentially counts the number of microstates that correspond to this macrostate. Note that for a system in equilibrium, the Boltzmann entropy is equal to the thermodynamic entropy.

Given that our coarse-graining cell size is $L_S$, the system consists of a single cell.   To compute the $S_B(t)$ for the system, we do the following steps - 
\begin{enumerate}
    \item Compute the state of the entire setup $\ket{\psi(t)}$ for all $t$ using time evolution block decimation (TEBD), and compute the average energy $E_{\rm sys}(t) = \expval{H_{\rm sys}}{\psi(t)}$ and the average magnetization $M_{\rm sys}(t) = \expval{S^z_{\rm sys}}{\psi(t)}$,  
    where $H_{\rm sys}$ [Eq.~\eqref{eq:system_hamiltonian}] and $S^z_{\rm sys}= \sum_{i = -(L_S-1)}^0S^z_i$ are the Hamiltonian and the total magnetization of the system.
    \item We next use the equivalence of ensembles and compute the entropy using the grand-canonical (GC) distribution 
    \begin{equation}
        \rho_{_{GC}}^{\rm sys}(t) = \frac{1}{Z(t)}e^{-\beta(t)(H_{\rm sys} - \mu(t) S^z_{\rm sys})},
        \label{eq:GCE}
    \end{equation}
    where $Z(t) = \Tr[e^{-\beta(t)(H_{\rm sys} - \mu(t) S^z_{\rm sys})}]$ is the grand-canonical partition function at time $t$. This involves finding $\beta$ and $\mu$ for all time steps such that the following two equations are satisfied:
    \begin{eqnarray}
        \label{eq:beta_mu1}
        \Tr\left[H_{\rm sys}\rho_{_{GC}}^{\rm sys}(t)\right] &=& E_{\rm sys}(t),\\
        \Tr\left[S^z_{\rm sys}\rho_{_{GC}}^{\rm sys}(t)\right] &=& M_{\rm sys}(t).
        \label{eq:beta_mu2}
    \end{eqnarray}
    This is done numerically by computing the roots [$\beta^*(t),\mu^*(t)$] of the above equations using Newton-Raphson (NR). The NR iterations are run till the expectation values obtained from the GC density matrix agree with the expectation values obtained using time dynamics up to some tolerance. For some cases, the NR method fails due to steep slopes in the functions described by the LHS of the above equations. In such cases, we solve these functions on a grid and search for solutions. The grid spacing is determined by the tolerance.
    \item Finally, using the $\beta^*(t),\mu^*(t)$, we construct the GC density matrix [Eq.~\eqref{eq:GCE}] and compute the corresponding Boltzmann entropy
    \begin{equation}
        S_{B}^{\rm sys} = -\Tr \rho_{_{GC}}^{\rm sys}\ln \rho_{_{GC}}^{\rm sys}.
        \label{eq:Sth}
    \end{equation}
\end{enumerate}

It must be noted that our system is not necessarily in local equilibrium.
The construction of the grand-canonical density matrix is done such that only the first moments of the conserved quantities are reproduced. However, for a truly thermal state, all the higher moments ought to be satisfied. Since our system is not truly in a thermal state, we do not expect to have such agreements for our system. However, we note that the GC is here being used simply as a mathematical tool to compute the Boltzmann entropy, \emph{i.e}, count the number of microstates consistent with the observed energy and particle number.

Unlike the $\svn$ for a pure state, the Boltzmann entropy of the system and bath are different. The Boltzmann entropy can also be computed for the bath with the following steps:
\begin{enumerate}
    \item As mentioned above, divide the bath into spatial bins where each bin has an equal number of sites (equal to the number of sites in the system).
    \item Compute the total energy $E_{\rm bin}(t) = \expval{H_{\rm bin}}{\psi(t)}$ and the total magnetization for each bin $M_{\rm bin}(t) = \expval{S^z_{\rm bin}}{\psi(t)}$ as a function of time. The Hamiltonian and the total magnetization of a bin are given by
    \begin{align}
        H_{\rm bin} &= J\sum_{i \in \text{bin}}\left(S^x_{i}S^x_{i+1} + S^y_{i}S^y_{i+1} + \Delta S^z_{i}S^z_{i+1}\right),\\
        S^z_{\rm bin} &= \sum_{i \in \text{bin}}S^z_i.        
    \end{align}
    Note that we have ignored the bond Hamiltonian between two bins.
    \item Compute $S_B$ for each bin using steps 2-3 of computing system $S_B$ mentioned above, treating each bin as the system and replacing $E_{\rm sys}$ and $M_{\rm sys}$ by $E_{\rm bin}$ and $M_{\rm bin}$ respectively.
    \item The total Boltzmann entropy of the bath is given by the sum of the Boltzmann entropy of each bin
    \begin{equation}
        S^{\rm bath}_{B} = \sum_{\rm bins}S^{\rm bin}_{B}.
        \label{eq:Sth_bath}
    \end{equation}
\end{enumerate}

\section{Early time regime}
\label{sec:early_time_analytics}

In this section, we provide analytical results for the growth of entanglement entropy and number statistics in the early time regime, up to $t\sim O(1/J)$. Since we are looking at very early times, the dynamics can be approximated by taking the first few terms of the Taylor expansion of the evolution operator
\begin{equation}
    e^{-\ii H t}\ket{\psi (0)} = \left[\mathbb{I} - \ii \, t H - \frac{t^2}{2}H^2 + O(t^3)\right]\ket{\psi (0)}.
    \label{eq:approx_dynamics}
\end{equation}
These terms considered in the Taylor expansion correspond to the timescale in which the first particle escapes from the system to the bath and the interactions do not play any role up to the order considered in Eq.~\eqref{eq:approx_dynamics}.

In the following, we first consider the case when the system is integrable ($J' = 0$) and then the non-integrable ($J' \neq 0$) case. 

\subsection{Integrable case $J' = 0$}
\label{sec:early_time_integrable}
We start with the polarized initial state [Eq.~\eqref{eq:sys_pol_init}], where the system is filled, that is, $\expval{N_{\rm sys}(0)} = \expval{\sum_{i \in \text{system}}(S^z_i +1/2)} = L_S$. At $t = 0$, the state of the full chain is given by $\ket{\psi(0)} = \ket{\psi_{\rm sys}(0)}\otimes \ket{\psi_{\rm bath}(0)}$, where $\ket{\psi_{\rm bath}(0)}$ is given by Eq.~\eqref{eq:micro_bath_pol_init}. From Eq.~\eqref{eq:approx_dynamics}, one can verify that the only sites that are affected by this dynamics are the 2 sites on either side of the system bath boundary. Therefore, the rest of the system and the bath remain decoupled from these 4 sites, and any dynamics is governed by the dynamics of these 4 sites. This statement follows from the observation that
\begin{align}
    \label{eq:H_psi}
    H\ket{\psi(0)} &= E_0 \ket{\psi(0)} + \frac{1}{2}\ket{\up}^{\otimes L_S-2}\otimes\ket{\psi_1}\otimes\ket{\dn}^{\otimes L_B-2}, \\
    H^2\ket{\psi(0)} &= \left(\frac{1}{4}+E_0^2\right)\ket{\psi(0)} +\ket{\up}^{\otimes L_S-2}\otimes\bigg[ \frac{E_0+E_1}{2}\ket{\psi_1} \nonumber \\
    & \hspace{2cm}+ \frac{1}{4}\left(\ket{\psi_2 } + \ket{\psi_3}\right)\bigg]\otimes\ket{\dn}^{\otimes L_B-2},
    \label{eq:H2_psi}
\end{align}
where $E_0 = (L-3)\Delta/4$ and $E_1= (L-7)\Delta/4$ and
\begin{eqnarray}
\label{eq:psi_def}
\ket{\psi_0} &=& \ket{\up\up\dn\dn},\quad \ket{\psi_1} = \ket{\up\dn\up\dn} \nonumber \\\ket{\psi_2} &=& \ket{\up\dn\dn\up},\quad \ket{\psi_3} = \ket{\dn\up\up\dn}\, .
\end{eqnarray}
We define the symbol, $\ket{x}^{\otimes M} \equiv \prod_{i=1}^M \otimes \ket{x}_i$ where $x$ can represent $\up,\dn$ and $M$ represents the number of sites under consideration in the system or the bath.  Using Eqs~\eqref{eq:H_psi} and \eqref{eq:H2_psi}, one can write the density matrix for the entire chain (system + bath) in terms of the 4-site states as [up to $O(t^2)$]
\begin{align}
    \rho(t) &= \Omega_S \otimes \Bigg\{\xi_{00} + \frac{\ii t}{2}\left[\xi_{01} - \xi_{01}^\dagger\right] - t^2\bigg[\frac{1}{4}\left(\xi_{00}-\xi_{11}\right) \nonumber\\ 
    & - \frac{\Delta}{4}\left(\xi_{01} + \xi_{01}^\dagger\right) + \frac{1}{8}\left(\xi_{02} + \xi_{02}^\dagger + \xi_{03} + \xi_{03}^\dagger\right) \bigg] \Bigg\}\otimes\Omega_B,
    \label{eq:rhot_early_time}
\end{align}
where 
\begin{eqnarray}
\label{eq:Omega}
\Omega_{S} = \ket{\up}^{\otimes L_{S}-2}\bra{\up}^{\otimes L_{S}-2}, \quad 
\Omega_{B} = \ket{\dn}^{\otimes L_{B}-2}\bra{\dn}^{\otimes L_{B}-2}\,  
\end{eqnarray}
and 
\begin{eqnarray}
\label{eq:xis}
\xi_{00} &=& \ket{\psi_0}\bra{\psi_0}, \quad\xi_{01} = \ket{\psi_0}\bra{\psi_1}, \quad \xi_{02} = \ket{\psi_0}\bra{\psi_2},\nonumber \\ \xi_{03} &=& \ket{\psi_0}\bra{\psi_3},\quad \xi_{11} = \ket{\psi_1}\bra{\psi_{1}}. 
\end{eqnarray}
Integrating out the bath in Eq.~\eqref{eq:rhot_early_time} is equivalent to integrating out the two bath sites from the $16\times16$ density matrix of the $4$ sites near the system-bath bond. The reduced-density matrix $\rho_S(t)$, in the basis \{$\ket{\up\up},\ket{\up\dn},\ket{\dn\up},\ket{\dn\dn} $, gets contribution only from the terms $\xi_{00}$ and $\xi_{11}$ in Eq.~\eqref{eq:rhot_early_time} and we get
\begin{align}
    \rho_S(t) = \Omega_S\otimes\begin{pmatrix}
1-\frac{t^2}{4} & 0 & 0&0\\
0 & \frac{t^2}{4} & 0&0\\
0 & 0 & 0&0\\
0 & 0 & 0&0 \, .
\end{pmatrix}
\label{eq:red_rho}
\end{align}
We finally find the entanglement [Eq.~\eqref{eq:SvN}]
\begin{align}
    \svn = -\left(1-\frac{t^2}{4}\right)\log\left(1-\frac{t^2}{4}\right) - \frac{t^2}{4}\log\frac{t^2}{4},
\end{align}
which on expanding gives, to $O(t^2)$,
\begin{equation}
    \svn = - \frac{t^2}{2}\log\frac{t}{2} + \frac{t^2}{4}.
    \label{eq:svn_early_time}
\end{equation}
As expected, up to $t \sim O(1/J)$, $\svn(t)$ is independent of $\Delta$. 

Similarly, one can also compute the average particle number in the  bath
\begin{align}
    \expval{N_{\rm bath}} = \expval{\sum_{i \in \text{bath}}\left(S^z_i +1/2\right)},
    \label{eq:Nbath_def}
\end{align} 
and its fluctuations, $var(N_{\rm bath})$, in this time regime. We notice that $N_{\rm bath}\ket{\psi (0)} = 0$, since at $t = 0$ the bath is empty, leading to
\begin{align}
    \label{eq:exp_Nbath_init}
    \expval{N_{\rm bath}} &= t^2\expval{HN_{\rm bath}H}{\psi(0)},\\
    var(N_{\rm bath}) &= t^2\expval{HN^2_{\rm bath}H}{\psi(0)} - \expval{N_{\rm bath}}^2,
    \label{eq:var_Nbath_init}
\end{align}
where we have used Eq.~\eqref{eq:approx_dynamics} to evolve the state up to $O(t^2)$. Using Eq.~\eqref{eq:H_psi}, one can show that up to $O(t^2)$
\begin{align}
    \label{eq:Nbath_early_time}
    \expval{N_{\rm bath}} &= \frac{t^2}{4},\\
    var(N_{\rm bath}) &= \frac{t^2}{4}.
    \label{eq:var_Nbath_early_time}
\end{align}

\textit{High entropy initial state: } So far, we have discussed the polarized initial state which has zero Boltzmann entropy. We now discuss the case when the system is initialized in a high $S_B$ state. Computing $\svn$ when the system is initialized in the high $S_B$ state is a much more challenging problem. However, computing $\expval{N_{\rm bath}}$ and $var(N_{\rm bath})$ is still feasible. Let us start with the state
\begin{align}
    \ket{\phi(0)} = \frac{1}{\sqrt{\mathbb{N}}}\sum_{k = 1}^{\mathcal{N}}c_k \ket{\chi_k}\otimes\ket{\dn}^{\otimes L_B},
    \label{eq:inf_temp_state}
\end{align}
where $\mathbb{N} = \sqrt{\sum_{k=1}^{\mathcal{N}}\abs{c_k}^2}$ is the normalization factor and $\mathcal{N} = {}^{L_S}C_{L_S/2}$ is the dimension of the half-filled sector of the system, $\{c_k\}$ are complex numbers chosen from a normal distribution with mean zero and variance $1/2$ and $\{\ket{\chi_k}\}$ are the $S^z$ product basis states of the system in the half-filled sector. If we consider the density matrix for the state $\ket{\phi(0)}$ and take an average over the choice of $\{c_k\}$, we find that
\begin{align}
    \overline{\ket{\phi(0)}\bra{\phi(0)}} &= \sum_{k,k' = 1}^{\mathcal{N}}\overline{\left(\frac{c_kc_{k'}^*}{\mathbb{N}}\right)}\ket{\chi_k}\otimes\ket{\dn}^{\otimes L_B}\bra{\chi_{k'}}\otimes\bra{\dn}^{\otimes L_B}.
\end{align}
For $\mathcal{N}>>1$, we expect, ignoring $1/\sqrt{\mathcal{N}}$ corrections,
\begin{align}
    \overline{\ket{\phi(0)}\bra{\phi(0)}} &=\left(\frac{1}{\mathcal{N}}\sum_{k=1}^{\mathcal{N}}\ket{\chi_k}\bra{\chi_{k}}\right)\otimes\ket{\dn}^{\otimes L_B}\bra{\dn}^{\otimes L_B},\\
    &= \frac{\mathbb{I}}{\mathcal{N}}\otimes\ket{\dn}^{\otimes L_B}\bra{\dn}^{\otimes L_B},
\end{align}
where we have used the fact that $\overline{Re(c_k)Re(c_k')} = \delta_{k,k'}/2$, $\overline{Im(c_k)Im(c_k')} = \delta_{k,k'}/2$, and $\overline{Re(c_k)Im(c_k')} = 0$. Therefore, the density matrix that we get from the state $\ket{\phi(0)}$ after averaging is a state where the system is in the infinite temperature state, while the bath remains empty. For a particular choice of $\{c_k\}$ one might end up with a low entropy state. However, for a large enough system, this is highly unlikely. Therefore, the state in Eq.~\eqref{eq:inf_temp_state} is an infinite temperature state for all practical purposes when $L_S$ is large enough. Since the bath is still empty, $N_{\rm bath}\ket{\phi(0)} = 0$ and Eqs.~\eqref{eq:exp_Nbath_init} and \eqref{eq:var_Nbath_init} holds for $\ket{\phi(0)}$. Now, $N_{\rm bath}H\ket{\chi_k}\otimes\ket{\dn}^{\otimes L_B}$ is non-zero for only those $\ket{\chi_k}$ which have the last system site in the $\ket{\up}$ state. Let, us denote these states as $\ket{\Tilde{\chi}_k, \up}$. Furthermore, $H\ket{\Tilde{\chi}_k, \up}\otimes\ket{\dn}^{\otimes L_B}$ leads to a linear combination of states due to the hopping term in the Hamiltonian [Eq.~\eqref{eq:sys_micro_bath_hamil}]. However, the only state which has non-zero particles in the bath is $(1/2)\ket{\Tilde{\chi}_k, \dn}\otimes\ket{\up}\ket{\dn}^{\otimes L_B-1}$. Therefore,
\begin{align}
    \expval{N_{\rm bath}} &= t^2 \expval{HN_{\rm bath}H}{\phi(0)}\\
    &=\frac{t^2}{2\mathbb{N}}\sum_{k,k'}{}^{'} c_kc_{k'}^*\bra{\chi_{k'}}\bra{\dn}^{\otimes L_B}H \ket{\Tilde{\chi}_k, \dn}\ket{\up}\ket{\dn}^{\otimes L_B-1}
\end{align}
where the primed sum $\sum{}^{'}$ indicates the sum over states where the last site in the system is in the $\ket{\up}$ state. Moreover, the only terms from $\bra{\chi_{k'}}\otimes\bra{\dn}^{\otimes L_B}H$ that survives are $(1/2)\bra{\Tilde{\chi}_{k'},\up}\otimes\bra{\up}\bra{\dn}^{\otimes L_B-1}$ due to orthogonality, leading to
\begin{align}
    \expval{N_{\rm bath}}&= \frac{t^2}{4\mathbb{N}}\sum_{k}{}^{'}\abs{c_k}^2,
\end{align}
Now, since the probability of the last system site to be in the $\ket{\up}$ state is $1/2$, therefore $\overline{(1/\mathbb{N})\sum_{k}{}^{'}\abs{c_k}^2} = 1/2$, leading to
\begin{align}
    \expval{N_{\rm bath}} = \frac{t^2}{8},
    \label{eq:inf_Nbath_early_time}
\end{align}
up to $O(t^2)$. Using the same steps, one can also compute the particle fluctuations in the bath up to $O(t^2)$
\begin{align}
    var(N_{\rm bath}) = \frac{t^2}{8}.
    \label{eq:inf_var_Nbath_early_time}
\end{align}
Again, as expected, this is independent of $\Delta$.

\subsection{Non-integrable case $J' = 1$}
\label{sec:early_time_non_integrable}

Now, for the case when the system is non-integrable ($J' = 1$), the results presented in Sec.~\ref{sec:early_time_integrable} do not change. Due to the NNN interactions, the initial state energy for the polarized case, changes from $E_0$ to $E_0^{\rm NNN} = (L-3)\Delta/4 + (L_S-2)J'/4$ and the energy in state $\ket{\up}^{\otimes L_S-2}\ket{\psi_1} \ket{\up}^{\otimes L_S-2}$ changes from $E_1$ to $E_1^{\rm NNN} = (L-7)\Delta/4 + (L_S-4)J'/4$. However, as pointed out previously (Sec.~\ref{sec:early_time_integrable}), the interactions does not play a role in this regime. Since the integrability breaking term is just NNN interactions, they do not vary the solutions in Eqs.~\eqref{eq:svn_early_time}, \eqref{eq:Nbath_early_time}, and \eqref{eq:var_Nbath_early_time}.

Similarly, when we start from the high $S_B$ initial state [Eq.~\eqref{eq:inf_temp_state}], the interactions do not play any role and Eqs.~\eqref{eq:inf_Nbath_early_time} and \eqref{eq:inf_var_Nbath_early_time} remain the same, up to $O(t^2)$, in this regime.

\section{Numerical results}
\label{sec:numerical_results}
In this section, we discuss the numerical results. To study the dynamical properties of the system, we use the time evolution block decimation (TEBD) method, where we represent states as matrix product states (MPS) \cite{USJan2011} using the ITensor library \cite{itensor}. We provide a comprehensive description of the method in Appendix~\ref{app:TEBD}. We consider the entire bath and the system as a closed chain whose dynamics are governed by the Hamiltonian in Eq.~\eqref{eq:sys_micro_bath_hamil}. For all TEBD simulations, we fix the bond dimension cutoff to $\chi = 150$, and the Trotter step to $\delta t = 0.05$ (in units of $1/J$). In Secs.~\ref{sec:integrable_numerics} and \ref{sec:non_integrable_numerics} we report the results for the cases where the system is integrable ($J' = 0$) and non-integrable ($J' = 1$), respectively. However, in all cases, the bath remains integrable.
In Table~\ref{tab:summary_table}, we summarize the power-laws of the $S_{vN}$, $var(N_{\rm bath})$ and $\expval{N_{\rm bath}}$ in the intermediate time scales, for the different cases discussed in Secs.~\ref{sec:integrable_numerics} and \ref{sec:non_integrable_numerics}.
\begin{table*}[t]
    \centering
    \begin{tabular}{|c|c|c|c|c|}
        \hline
        \multirow{2}{*}{\qquad Quantity\qquad}&\multicolumn{2}{c|}{$\Delta = 0.8J$}&\multicolumn{2}{c|}{$\Delta = 1.0J$}\\
        \cline{2-5}
        &$\qquad J' = 0 \qquad$ &$\qquad J' = 1\qquad$&$\qquad J' = 0 \qquad$&$\qquad J' = 1 \qquad$ \\        
        \hline
        \multicolumn{5}{|c|}{Filled initial state}\\ \hline
        {$S_{vN}$}& $0.3 \pm 0.004$ &  & $0.307 \pm 0.002$ &  \\
        {$var(N_{\rm bath})$}& $0.507 \pm 0.012$ & N.A. & $0.584 \pm 0.007$ & N.A. \\
        {$\expval{N_{\rm bath}}$}& $0.826 \pm 0.002$ &  & $0.658 \pm 0.005$ &  \\
        \hline
        \multicolumn{5}{|c|}{Infinite temperature initial state}\\
        \hline
        {$S_{vN}$}& $0.804 \pm 0.005$ & $0.79 \pm 0.003$ & $0.777 \pm 0.003$ & $0.767 \pm 0.002$ \\
        {$var(N_{\rm bath})$}& $0.726 \pm 0.003$ & $0.697 \pm 0.003$ & $0.649 \pm 0.004$ & $0.66 \pm 0.002$ \\
        {$\expval{N_{\rm bath}}$}& $0.932 \pm 0.003$ & $0.801 \pm 0.004$ & $0.843 \pm 0.003$ & $0.758 \pm 0.003$ \\
        \hline
    \end{tabular}
    \caption{Summary of results for the growth of entanglement in the different cases when the system is connected to an interacting integrable bath, in the intermediate time regime [from $t \sim O(1/J)$ to $t_{\rm Page}$]. The exponent $\alpha$ of the scaling $t^\alpha$ is reported. N.A. in the table implies those cases where the dynamics freezes, thereby lacking a Page curve.}
    \label{tab:summary_table}
\end{table*}

\begin{figure*}[t]
    \centering
    \includegraphics[width=\linewidth]{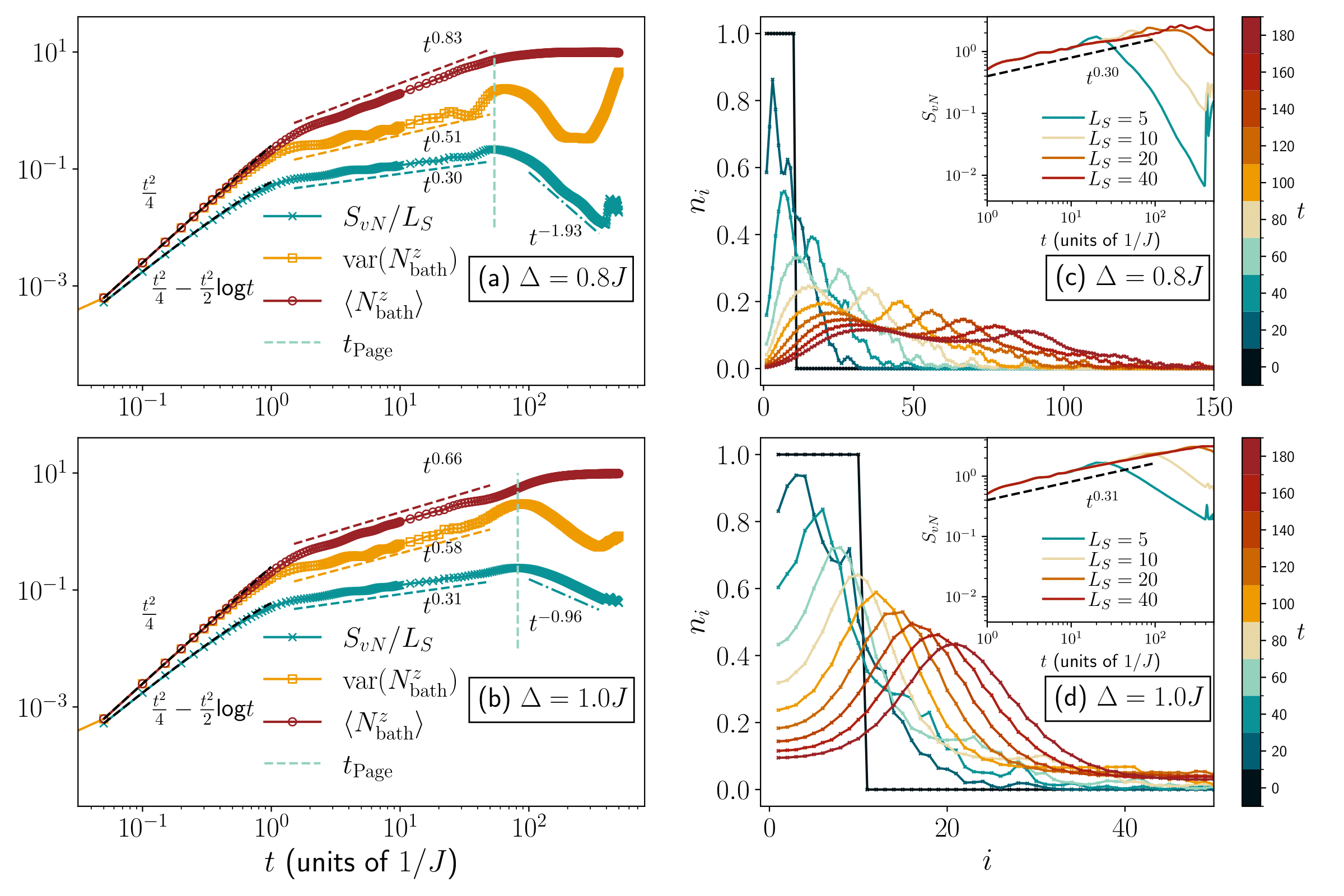}
    \caption{The entanglement entropy $S_{vN}(t)$, the mean $\expval{N_{\rm bath}}$, and the fluctuation $var(N_{\rm bath})$ of the particle number of the bath, for the integrable case ($J' = 0$), are plotted as a function of time for (a) $\Delta = 0.8J$ and (b) $\Delta = 1.0J$ with $L_S = 10$ and $L_B = 200$. We start from a filled state, and the bath is interacting. The early time data [$t$ up to $O(1/J)$] shows excellent agreement with the analytical solutions given by Eq.~\eqref{eq:svn_early_time} for $\svn$ and Eqs.~\eqref{eq:Nbath_early_time} and \eqref{eq:var_Nbath_early_time} for $\expval{N_{\rm bath}}$ and $var(N_{\rm bath})$ respectively. In (c) and (d), we plot the density profile $n_i$, for different time instances, for $\Delta = 0.8J$ and $1.0J$, respectively. In the insets of (c) and (d), we plot the $S_{vN}(t)$ in the intermediate regime for $L_S = 5,10,20$ and $40$, keeping $L_B = 200$ for $\Delta = 0.8J$ and $1.0J$ respectively.}
    \label{fig:Pol_dynamics}
\end{figure*}

\begin{figure*}[t]
    \centering
    \includegraphics[width=1.0\linewidth]{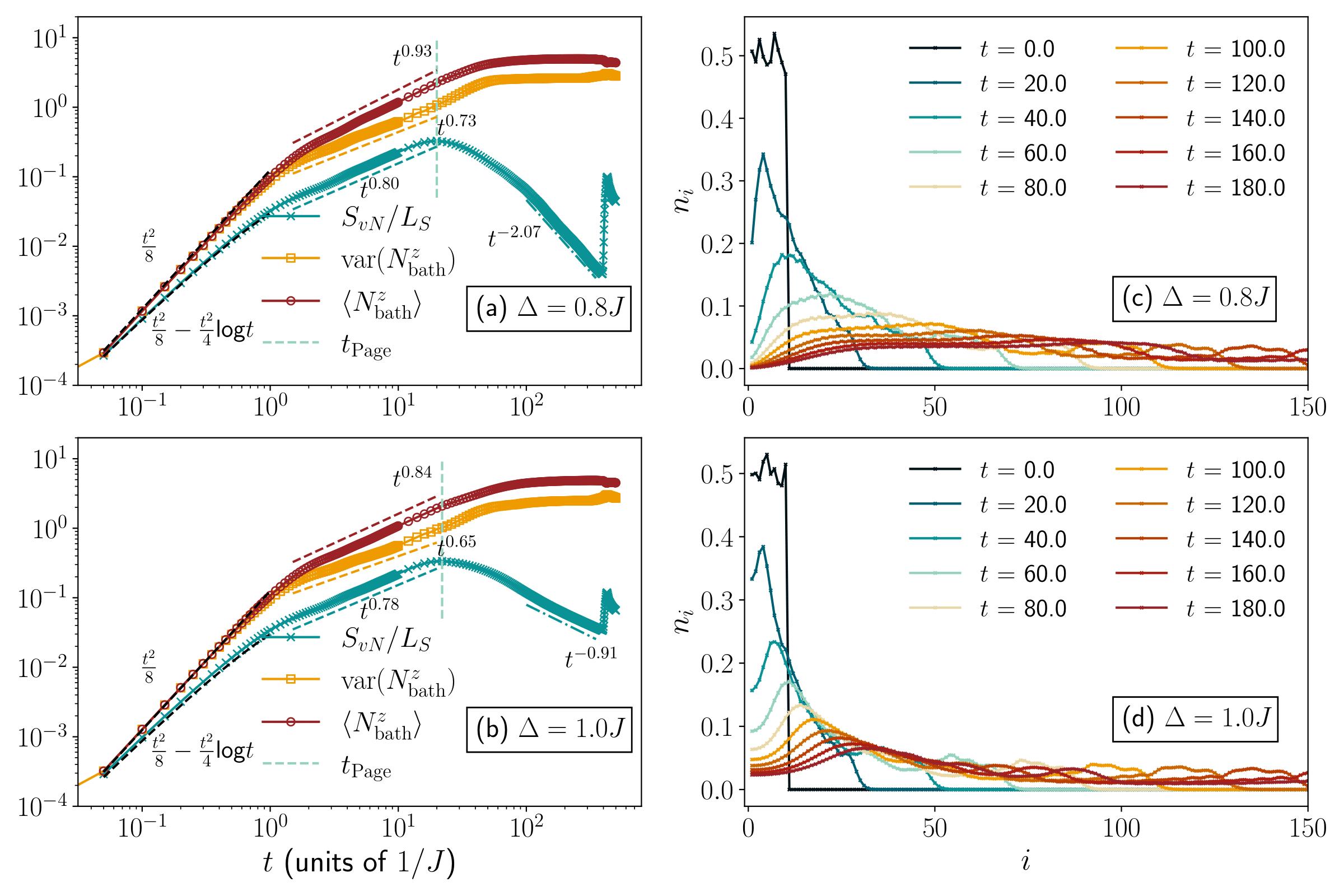}
    \caption{$S_{vN}(t)$, $\expval{N_{\rm bath}}$, and $var(N_{\rm bath})$ of the particle number of the bath, for the integrable case ($J' = 0$), are plotted as a function of time for (a) $\Delta = 0.8J$ and (b) $\Delta = 1.0J$ with $L_S = 10$ and $L_B = 200$. We start from a high entropy state and the bath is interacting. The early time data [$t$ up to $O(1/J)$] for $\expval{N_{\rm bath}}$ and $var(N_{\rm bath})$, shows excellent agreement with the analytical solutions given by Eqs.~\eqref{eq:inf_Nbath_early_time} and \eqref{eq:inf_var_Nbath_early_time} respectively. However, for $\svn$, we do not have an analytical expression. Nonetheless, the expression $t^2/8 - (t^2/4)\log t$ seems to be a good fit. In (c) and (d), we plot the density profile $n_i$, for different time instances, for $\Delta = 0.8J$ and $1.0J$ respectively.}
    \label{fig:high_temp_dynamics}
\end{figure*}

\begin{figure*}[t]
    \centering
    \includegraphics[width=1.0\linewidth]{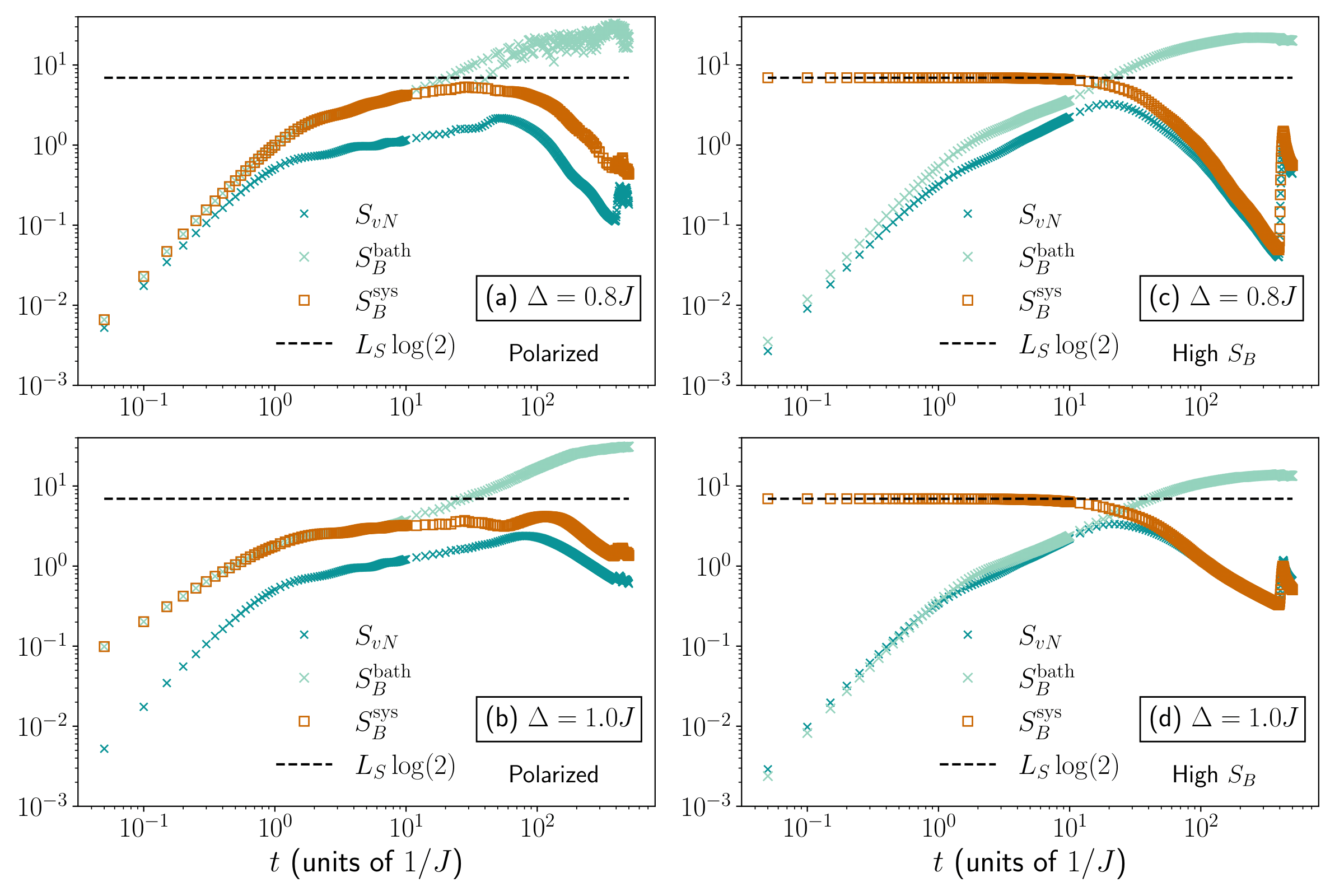}
    \caption{$S_{vN}(t)$ and $S_B(t)$ of the system and bath for the integrable case ($J' = 0$), are plotted as a function of time for the polarized initial condition (a,b) and for the high entropy case (c,d) for the two values of  $\Delta = 0.8J$ and $\Delta = 1.0J$.  We consider $L_S = 10$ and $L_B = 200$, and the bath is interacting.}
    \label{fig:Sth_dynamics}
\end{figure*}

\subsection{Integrable case $J' = 0$}
\label{sec:integrable_numerics}

\textit{Filled (spin-polarized) initial state:} We start with the polarized initial state [Eq.~\eqref{eq:sys_pol_init}], where the system is filled. As stated in the previous section, $\expval{N_{\rm sys}(0)} = L_S$. At $t = 0$, the state of the full chain is given by $\ket{\psi(0)} = \ket{\psi_{\rm sys}(0)}\otimes \ket{\psi_{\rm bath}(0)}$, where $\ket{\psi_{\rm bath}(0)}$ is given by Eq.~\eqref{eq:micro_bath_pol_init}. In Fig.~\ref{fig:Pol_dynamics}, we plot the $S_{vN}(t)$, the mean $\expval{N_{\rm bath}}$, and the fluctuation $var(N_{\rm bath})$ of the particle number of the bath, for the integrable case ($J' = 0$), as a function of time for (a) $\Delta = 0.8J$ and (b) $\Delta = 1.0J$ with $L_S = 10$ and $L_B = 200$. We find three prominent time regimes in the dynamics. In the first regime, up to $t\sim O(1/J)$, the numerics agree with the analytical solutions obtained in Eqs.~\eqref{eq:svn_early_time}, \eqref{eq:Nbath_early_time}, and \eqref{eq:var_Nbath_early_time}. In the intermediate regime, for $\Delta = 0.8J$ the three quantities grow as 
\begin{align}
    \svn \sim t^{0.30},\, \expval{N_{\rm bath}} \sim t^{0.83}, \, var(N_{\rm bath}) \sim t^{0.51},
    \label{eq:int_regime_results1}
\end{align}
while for $\Delta = 1.0J$
\begin{align}
    \svn \sim t^{0.31},\, \expval{N_{\rm bath}} \sim t^{0.66}, \, var(N_{\rm bath}) \sim t^{0.58}.
    \label{eq:int_regime_results2}
\end{align}
 These exponents for $\svn$ are different from the one reported in Refs.~\onlinecite{GMNov2017,ML2017}, where the authors report that the entanglement growth exponent is approximately $0.25$. This difference in exponents is due to the fact that these studies consider infinite domain walls, while in our case, the system is finite. In fact, the faster growth of entanglement in our case can be attributed to the reflection of the spin current from the left edge (see Fig.~\ref{fig:schematic}), which further enhances the system-bath entanglement.  
Note that, for $\Delta = 0$ (non-interacting case), both $\svn$ and the $var(N_{\rm bath})$ grow logarithmically in the intermediate regime \cite{OG2020}. This is not the case for the interacting case. 

The intermediate regime entanglement growth eventually stops at the Page time, $t_{\rm Page}$. We find from our finite size studies [see insets of Figs.~\ref{fig:Pol_dynamics} (c) and (d)] that $t_{\rm Page} \approx L_S^{1.7}$ for $\Delta =0.8$ and $t_{\rm Page} \approx L_S^2$ for $\Delta =1.0$. Moreover, at these Page times, we find that approximately $0.74$ and $0.55$ fraction of the particles has escaped from the system for $\Delta = 0.8J$ and $\Delta = 1.0J$, respectively. After the Page time, for $\Delta = 0.8J$, the $\svn$ falls of as $1/t^{1.93}$, and for $\Delta = 1.0J$, the $\svn$ falls off as $1/t^{0.96}$. In Fig.~\ref{fig:Pol_dynamics} (c) and (d), we plot the density profile $n_i = \expval{S^z_i +1/2}$, for different time instances, for $\Delta = 0.8J$ and $1.0J$ respectively. As expected from the growth of $N_{\rm bath}$, the particles dissipate faster into the bath for $\Delta = 0.8J$ than for $\Delta = 1.0J$. This is consistent with the fact that particles propagate ballistically for $\Delta<1.0J$ and super-diffusively for $\Delta = 1.0J$ when the system is initialized in the domain wall state with $L_S = L_B$ \cite{ML2017}. In the insets of Fig.~\ref{fig:Pol_dynamics} (c) and (d), we plot the $S_{vN}(t)$ in the intermediate regime for $L_S = 5,10,20$ and $40$, keeping $L_B = 200$ for $\Delta = 0.8J$ and $1.0J$ respectively. The intermediate regime increases with increasing $L_S$, confirming the power-law behavior of the entanglement.

\textit{High entropy initial state:} It must be noted that the initial state of the system considered above has zero Boltzmann entropy. While the dynamics of entanglement from such a quench is interesting, it is also essential to investigate the fate of entanglement when the system is initialized in a high entropy state. Moreover, such a setup has more resemblance to the initial state of a black hole \cite{DPDec1993,DPSep2013}. To this end, we prepare the system in an infinite temperature random state. Note that, preparing the MPS for a random state described in Eq.~\eqref{eq:inf_temp_state} is an extremely difficult task for large baths. However, one can start with the MPS of a simple product state and apply Haar random unitary gates on this state such that the resulting state is very close to a state with high Boltzmann entropy and is close to half-filling. Although such a state is not exactly the same as the one mentioned in Eq.~\eqref{eq:inf_temp_state}, we expect the analytical results obtained in Sec.~\ref{sec:early_time_analytics} to hold. The details of preparing the MPS for such a state are provided in Appendix~\ref{app:TEBD}. The Boltzmann entropy $S_B$, as defined in Sec.~\ref{sec:entropy}, for such a state is close to that of the infinite temperature state and $S_{B}(0) \approx L_S\log 2$. In Fig.~\ref{fig:high_temp_dynamics}, we plot  $S_{vN}(t)$, $\expval{N_{\rm bath}}$ and $var(N_{\rm bath})$  for the integrable case ($J' = 0$), as a function of time for (a) $\Delta = 0.8J$ and (b) $\Delta = 1.0J$ with $L_S = 10$ and $L_B = 200$, starting from a high entropy state, for the system, and an empty bath. In the first regime [up to $t \sim O(1/J)$], we find that the dynamics of these quantities have similar growth laws, but with different pre-factors,  compared to the case when the system is in the zero entropy filled state [Sec.~\ref{sec:early_time_integrable}]. In particular, $\expval{N_{\rm bath}}$ and $var(N_{\rm bath})$ in this time regime are given by Eq.~\eqref{eq:inf_Nbath_early_time} and \eqref{eq:inf_var_Nbath_early_time} respectively. In the intermediate regime, for $\Delta = 0.8J$, the above three quantities grow as
\begin{align}
    \svn \sim t^{0.80},\, \expval{N_{\rm bath}} \sim t^{0.93}, \, var(N_{\rm bath}) \sim t^{0.73},
    \label{eq:heNN8}
\end{align}
while for $\Delta = 1.0J$
\begin{align}
    \svn \sim t^{0.78},\, \expval{N_{\rm bath}} \sim t^{0.84}, \, var(N_{\rm bath}) \sim t^{0.65}.
    \label{eq:heNN1}
\end{align}
All the growth rates are faster than the case when the system is in the filled initial state [Eqs.~\eqref{eq:int_regime_results1} and \eqref{eq:int_regime_results2}]. 
This trend of faster decay of particles into the bath is consistent with the results reported in Ref.~\onlinecite{ML2017}, where the authors found that when the system is initialized close to an infinite temperature state, the domain wall broadens as $t^{2/3}$, which is faster compared to $t^{3/5}$ broadening observed when the system is polarized.  Consequently, the faster growth in the entanglement can be explained by the faster decay of particles from the system to the bath when it is initialized in a high $S_B$ state.
At the Page time, $0.45$ and $0.41$ fractions of the particles have decayed into the bath for $\Delta = 0.8J$ and $\Delta = 1.0J$, respectively.
Finally, after the Page time, the $\svn$ falls off as $1/t^{2.07}$ for $\Delta = 0.8J$ and $1/t^{0.91}$ for $\Delta = 1.0J$, which is similar to the case when the system is filled. In Fig.~\ref{fig:high_temp_dynamics} (c) and (d), we plot the density profile $n_i$, for different time instances, for $\Delta = 0.8J$ and $1.0J$, respectively. It must be noted that choosing the initial state in a manner [Appendix~\ref{app:TEBD}] such that the system has high Boltzmann entropy ($\sim L_S \log 2$) ensures that the system is close to the half-filled state. 

Finally, we discuss the comparison between the entanglement entropy,  $S_{vN}(t)$, and the Boltzmann entropy, $S_B(t)$. For the computation of $S_B(t)$, we follow the procedure described in Sec.~\ref{sec:entropy}. First, we consider the case when the system is initialized in the polarized state. In Fig.~\ref{fig:Sth_dynamics} (a) and (b), we plot  $\svn (t)$, $S_B^{\rm sys}(t)$, and $S_B^{\rm bath}(t)$ for the integrable case ($J' = 0$), as a function of time for $\Delta = 0.8J$ and $\Delta = 1.0J$ respectively. At $t = 0$, both the system and the bath are in a polarized state (the system in the filled state and the bath in the empty state) and have zero Boltzmann entropy. As the system dissipates particles into the bath, $S_B^{\rm sys}$ increases till it reaches a maximum value at a time close to the Page time,  $t_{\rm Page}$, after which it starts decreasing. Interestingly, $S_B^{\rm sys}$ has a similar form as the Page curve, though it is always larger than the entanglement entropy at any instant of time. The $S_B^{\rm bath}$ initially has the same form as the system but it continues to increase even beyond $t_{\rm Page}$, till it reaches its maximum possible value given that the bath and the number of particles are finite in our numerical computations.

Next, we discuss the evolution of $S_B (t)$ for the case when the system is initialized in the high entropy state. In Fig.~\ref{fig:Sth_dynamics} (c) and (d), we plot  $S_{vN}(t)$, $S_B^{\rm sys}(t)$, and $S_B^{\rm bath}(t)$, as a function of time for $\Delta = 0.8J$ and $\Delta = 1.0J$ respectively.  At $t = 0$, the system is in a high entropy pure state, which is almost half-filled, and $S_B^{\rm sys}\approx L_S\log 2$, whereas the bath is in an empty pure state, which has zero Boltzmann entropy. As the bath gets filled, $S_B^{\rm bath}$ increases. Interestingly, we find that this rise in entropy is close to the growth of the entanglement entropy up to the Page time. On the other hand, as the particles in the system decay into the bath, $S_B^{\rm sys}$ can only decrease. We find that after the Page time, the $S_B^{\rm sys}$ follows the entanglement entropy. Eventually, as all the particles leave the system, it is expected that both $S_{vN}(t)$ and 
$S_B^{\rm sys}(t)$ go to zero. 

\begin{figure}[h]
    \centering
    \includegraphics[width=1.0\linewidth]{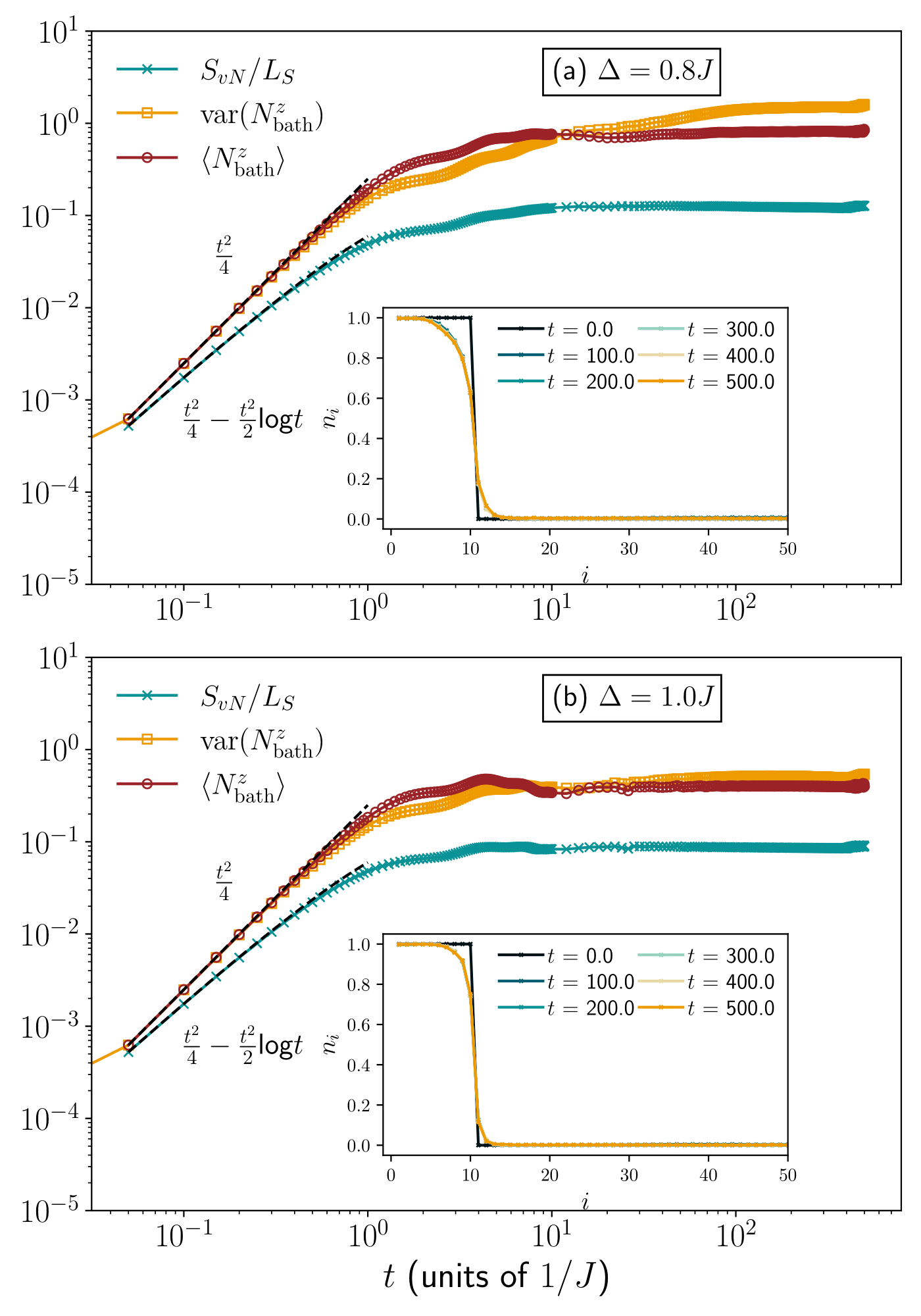}
    \caption{The entanglement entropy $S_{vN}(t)$, the mean $\expval{N_{\rm bath}}$, and the fluctuation $var(N_{\rm bath})$ of the particle number of the bath, for the non-integrable case with $J' = 1$, are plotted as a function of time for (a) $\Delta = 0.8J$ and (b) $\Delta = 1.0J$ with $L_S = 10$ and $L_B = 200$. We start from a filled state, and the bath is interacting. In the insets of (a) and (b), we plot the density profile for different time instances for $\lambda = 0.8J$ and $1.0J,$ respectively.}
    \label{fig:NNN_dynamics}
\end{figure}

\begin{figure*}[t]
    \centering  \includegraphics[width=1.0\linewidth]{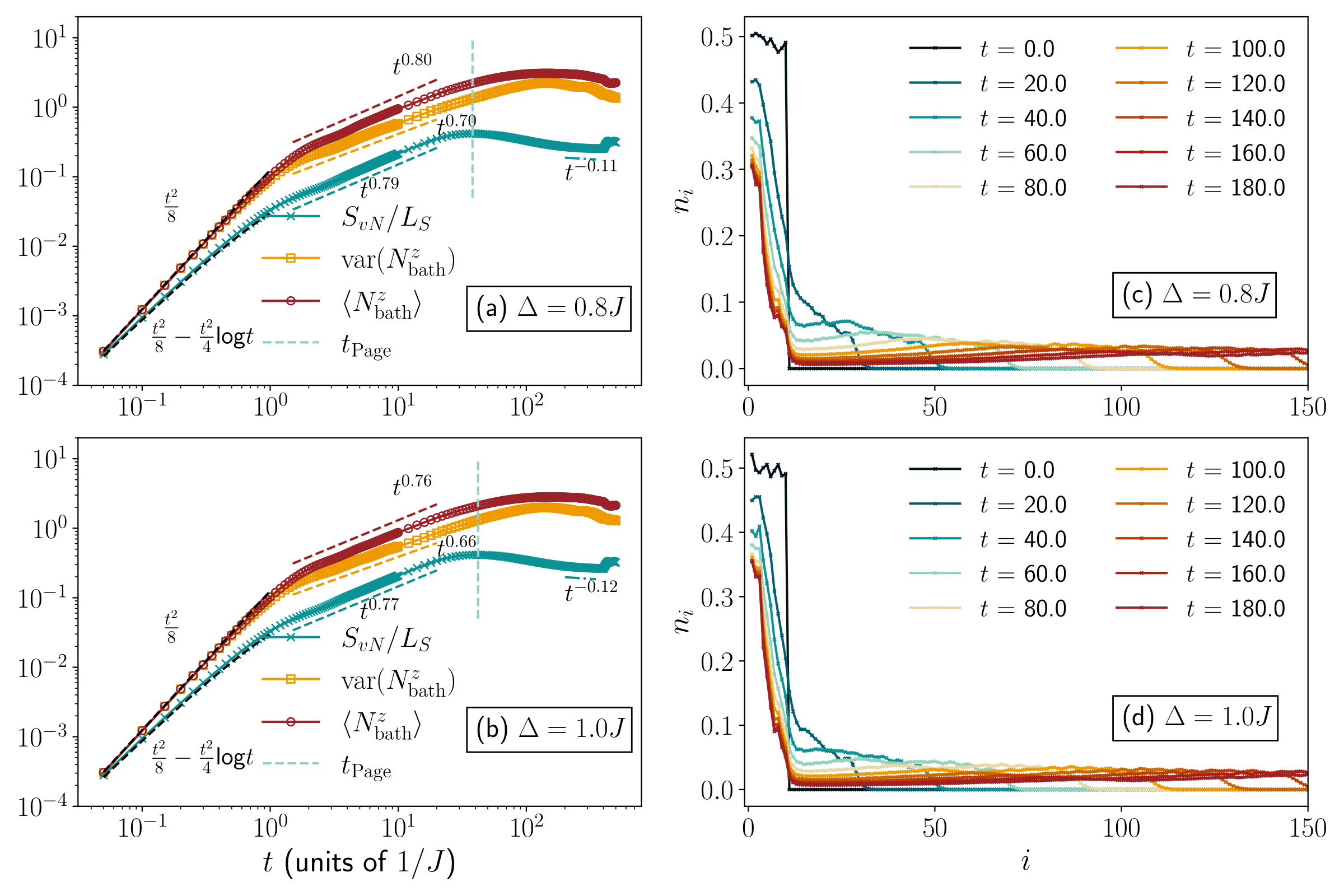}
    \caption{$S_{vN}(t)$, $\expval{N_{\rm bath}}$ and $var(N_{\rm bath})$ of the particle number of the bath, for the non-integrable case with $J' = 1$, are plotted as a function of time for (a) $\Delta = 0.8J$ and (b) $\Delta = 1.0J$ with $L_S = 10$ and $L_B = 200$. We start from a high-Boltzmann entropy state, and the bath is interacting. The early time data [$t$ up to $O(1/J)$] for $\expval{N_{\rm bath}}$ and $var(N_{\rm bath})$, matches exactly with the analytical solutions given by Eqs.~\eqref{eq:inf_Nbath_early_time} and \eqref{eq:inf_var_Nbath_early_time} respectively. However, for $\svn$, we do not have an analytical expression. Nonetheless, the expression $t^2/8 - (t^2/4)\log t$ seems to be a good fit. In (c) and (d), we plot the density profile $n_i$, for different time instances, for $\Delta = 0.8J$ and $1.0J$, respectively.}
    \label{fig:NNN_high_temp_dynamics}
\end{figure*}

\subsection{Non-integrable case $J' = 1$}
\label{sec:non_integrable_numerics}
In this section, we study the dynamics for the case where the system has non-integrable interactions and is connected to an integrable bath. We report results for two initial conditions.

\textit{Filled (spin-polarized) initial state:}  In Fig.~\ref{fig:NNN_dynamics}, we plot the time evolution of  $S_{vN}(t)$,  $\expval{N_{\rm bath}}$, and  $var(N_{\rm bath})$ for the case when the system is in a filled state at $t = 0$, for (a) $\Delta = 0.8J$ and (b) $\Delta = 1.0J$. We set  $L_S = 10$ and $L_B = 200$. As mentioned in Sec.~\ref{sec:early_time_non_integrable}, up to $t \lesssim O(1/J)$, the interactions do not play a role in the dynamics and we get the same growth form as in the case when the system is integrable ($J' = 0$). The numerical results for both $\Delta = 0.8J$ and $1.0J$ agree with the analytical solutions given by Eqs.~\eqref{eq:svn_early_time}, \eqref{eq:Nbath_early_time}, and \eqref{eq:var_Nbath_early_time}. 
After this initial growth,  $\svn$, $\expval{N_{\rm bath}}$, and $var(N_{\rm bath})$ all eventually seem to saturate to constant values up to the times investigated numerically, for both $\Delta = 0.8J$ and $\Delta = 1.0J$. In the insets of Fig.~\ref{fig:NNN_dynamics} (a) and (b), we plot the density profiles at different time instances, and we note that only a few particles ($\approx 1$) decay into the bath. The dynamics of the system freeze beyond this point. This can be attributed to the fact that the filled initial state has a high overlap with a single localized eigenstate of the entire setup (system and bath). We have numerically verified this by computing the full spectrum of the system plus bath Hamiltonian (see Appendix~\ref{app:overlap}). As a result, the dynamics basically freeze, and we do not see further growth in the entanglement.

\textit{High entropy initial state:} Similar to Sec.~\ref{sec:integrable_numerics}, we investigate the dynamics when the system is initialized in a high Boltzmann entropy state. In this case, we observe a Page-like behavior in the entanglement of the system. In Fig.~\ref{fig:NNN_high_temp_dynamics}, we plot $S_{vN}(t)$, $\expval{N_{\rm bath}}$, and $var(N_{\rm bath})$ for (a) $\Delta = 0.8J$ and $\Delta = 1.0J$ with $J' = 1$, $L_S = 10$ and $L_B = 200$. For the $\Delta = 0.8J$, the three quantities grow as
\begin{align}
    \svn \sim t^{0.79},\, \expval{N_{\rm bath}} \sim t^{0.80}, \, var(N_{\rm bath}) \sim t^{0.70},
    \label{eq:nnn8}
\end{align}
and for $\Delta = 1.0J$,
\begin{align}
    \svn \sim t^{0.77},\, \expval{N_{\rm bath}} \sim t^{0.76}, \, var(N_{\rm bath}) \sim t^{0.66}.
    \label{eq:nnn1}
\end{align}
In this case of high entropy initial state, we notice that this type of additional integrability-breaking interaction [Eq.~\eqref{eq:system_hamiltonian}] does not have much effect on the power-law scaling up to the Page time for the quantities. In other words, note that the exponents appearing in Eqs.~\eqref{eq:nnn8} and \eqref{eq:nnn1} for the nonintegrable case are close to those for the integrable case given in Eqs.~\eqref{eq:heNN8} and \eqref{eq:heNN1}. At the Page time, approximately $0.45$ and $0.42$ fraction of the particles have escaped to the bath for $\Delta = 0.8J$ and $\Delta = 1.0J$ respectively. After the Page time, the entanglement decays as $1/t^{0.11}$ for $\Delta = 0.8J$ and $1/t^{0.12}$ for $\Delta = 1.0J$, which is much slower compared to the integrable case when $J' = 0$. In Fig.~\ref{fig:NNN_high_temp_dynamics} (c) and (d), we plot the density profiles at different time instances. In this case, the particles decay into the bath, and eventually, the system loses about half the particles, and we do not see the freezing observed for the filled initial state. Correspondingly, we have also verified that in this case the initial state has an overlap with a broad band of extended eigenstates of the system plus bath Hamiltonian (see Appendix~\ref{app:overlap}).

\section{Conclusions and outlook}
\label{sec:conclusions}
To summarize, in this article, we investigated the quantum dynamics of the von-Neumann entanglement entropy $S_{vN}$ for a finite interacting system ($XXZ$ spin - $1/2$), with and without next-nearest neighbor integrability breaking interactions, connected to a much larger integrable interacting bath ($XXZ$ spin - $1/2$). We also computed the mean particle number $\expval{N_{\rm bath}}$ and the particle fluctuations $var(N_{\rm bath})$ in the bath at any time. We considered two different initial states for the system, a filled state with zero Boltzmann entropy and a high-Boltzmann entropy state. The bath is always taken to be empty. Such a setup is inspired by the paradigm of black-hole evaporation and the famous black-hole information paradox. Irrespective of this resemblance, it is in itself interesting to study the growth and decay of entanglement in interacting quantum systems. We use MPS techniques to numerically explore large bath sizes, which is otherwise not feasible with exact diagonalization. 

For the integrable case, we found that the entanglement exhibits Page curve like behavior for both initial conditions, that is, the entanglement grows up to some point in time (known as the Page time), after which it decays. This behavior is associated with the decay of particles from the system to the bath. However, for the non-integrable case, the zero-entropy and high entropy initial conditions show very different behaviour. We found that for the low-entropy initial state, the entanglement saturates to a constant value up to the times explored numerically. This is due to a large overlap of the initial state with a bound state, as a result of which the dynamics freezes. In contrast, when one starts from the high-Boltzmann entropy state,  the entanglement has a Page curve like behavior. The results for the growth of entropy are summarized in Table~\ref{tab:summary_table}.

Another focus of our work was to consider the dynamics of coarse-grained Boltzmann entropy, $S_B$, of the system and bath, and compare this with the fine-grained entanglement entropy. We first proposed a method to quantify the Boltzmann entropy, $S_B$, for the system and for the bath. When the system is initiated in the zero Boltzmann entropy state, we show that $S_B^{\rm sys}$ has the form of a Page curve but differs quantitatively. On the other hand, $S_B^{\rm bath}$  agrees with $S_{vN}$ till around the Page time, after which it continues to increase. Starting from a high Boltzmann entropy state, we show that $S_B^{\rm sys}$ decreases as the particles dissipate into the bath, coinciding with $S_{vN}$ \emph{after} the Page time. On the other hand, $S_B^{\rm bath}$ increases as the particles fill the bath and agrees with $S_{vN}$ \emph{before} the Page time. A clear understanding of the relation between entanglement entropy and Boltzmann entropy and their close agreement in certain time regimes, for either system or bath, is an interesting open question. We believe that the observed closeness could be due to the fact that the reduced density matrix (of either the system or the reservoir) is close to a local equilibrium state.

In Appendix~\ref{app:non_int_bath}, we discuss entanglement dynamics when the system is connected to a non-interacting bath. Such a case has been studied recently in Ref.~\onlinecite{RJFeb2025}. We find that the entanglement grows faster compared to the case when the system is connected to an interacting bath.

We believe that our study is relevant in the context of the black hole entanglement entropy problem and entanglement in interacting systems in general. This problem has been investigated using non-interacting Fermions \cite{SKJune2024,MSFeb2024}, where they obtain the Page curve behavior. A natural question is to ask how the entanglement dynamics change when the system has interactions and what role integrability plays in the dynamics. 
 For non-interacting systems in the absence of a defect between the system and the bath, $S_{vN}$ is known to grow logarithmically~\cite{PCOct2007} for an infinite domain wall initial condition. In contrast, we find that the interactions enhances the growth of entanglement, and we get sub-linear power law growth.  Furthermore, for non-interacting systems, the presence of a defect between the system and the bath leads to linear growth in entanglement \cite{PCApr2005,MSFeb2024}. However, for the interacting case, we do not need any defect to achieve power-law behavior.

Our study thus provides a step towards understanding the entanglement dynamics in interacting systems. In the future, it would be interesting to study the entanglement dynamics using the hydrodynamic approach for integrable systems. In Ref.~\onlinecite{MSFeb2024}, the authors have shown an agreement between the hydrodynamic Yang-Yang entropy \cite{VANov2021} and the entanglement entropy for non-interacting Fermionic system. The hydrodynamic equations for the Heisenberg model are known \cite{BBNov2016,JNOct2018} and the study of the Yang-Yang entropy for the case discussed in this article can lead to a better understanding of the entanglement dynamics. In Ref.~\onlinecite{HLFeb2025}, the authors have provided a hydrodynamic interpretation of the entanglement dynamics for interacting systems which is similar to the Boltzmann entropy discussed in this article. However, a complete hydrodynamic study of the entanglement dynamics for interacting systems is still lacking. 

\textit{Note added: } During the finalization of the work, Refs.~\onlinecite{RJFeb2025,HLFeb2025} appeared on related topics. However, specific questions investigated are distinct.

\section*{Acknowledgements}
We gratefully acknowledge David Huse for many important discussions and suggestions during the course of this work. AD and MK thank the VAJRA faculty scheme (No. VJR/2019/000079) from the Science and Engineering Research Board (SERB), Department of Science and Technology, Government of India. AD acknowledges the J.C. Bose Fellowship
(JCB/2022/000014) of the Science and Engineering Research Board of the Department of Science and Technology, Government of India. We acknowledge support from the Department of Atomic Energy, Government of India, under project No.~RTI4001.

\appendix

\setcounter{figure}{0}
\renewcommand{\thefigure}{A\arabic{figure}}

\section{System coupled to non-interacting bath}
\label{app:non_int_bath}

\begin{figure*}[t]
    \centering
    \includegraphics[width=1.0\linewidth]{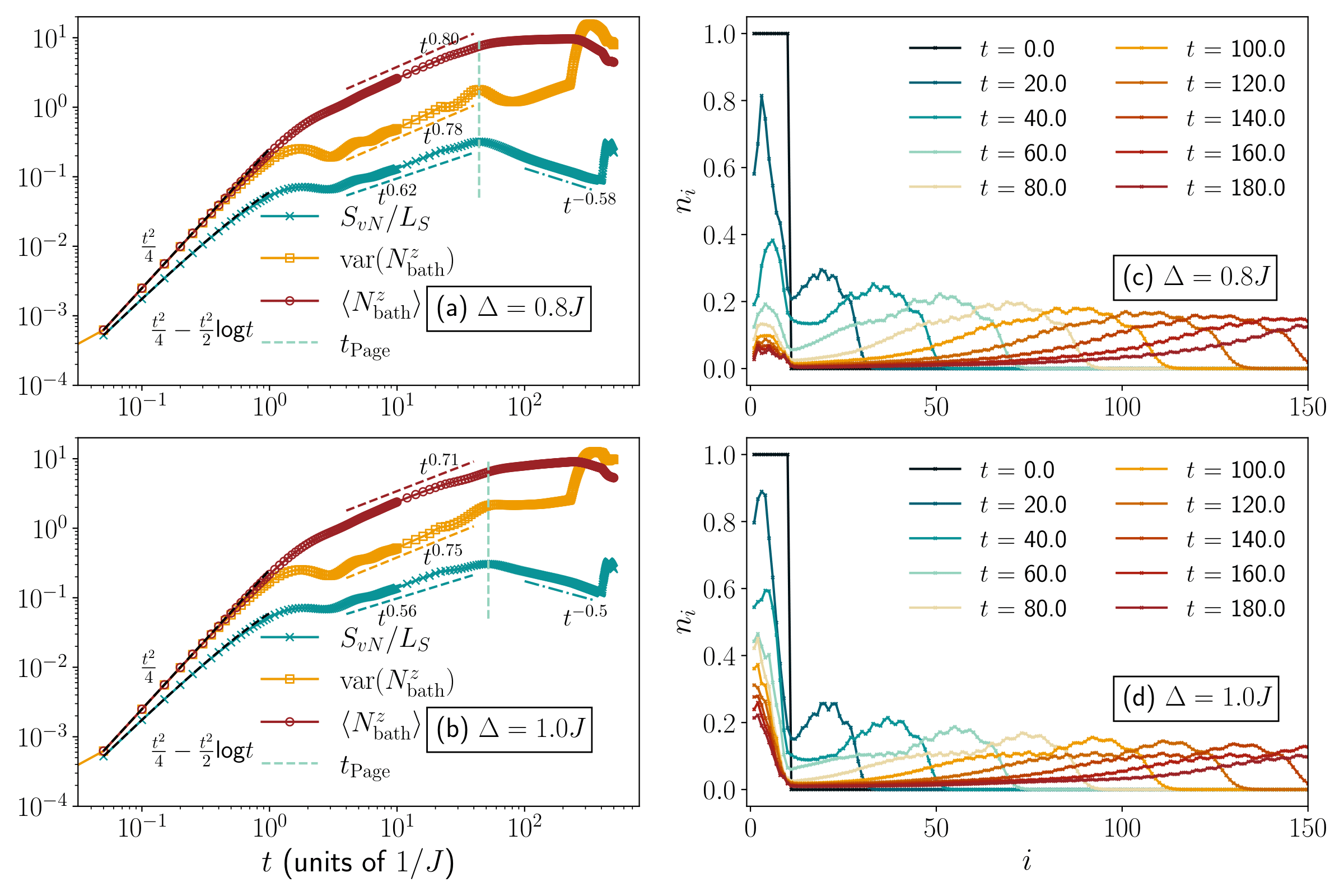}
    \caption{The $S_{vN}(t)$, the mean $\expval{N_{\rm bath}}$, and the fluctuation $var(N_{\rm bath})$ of the particle number of the bath, for an integrable interacting system, are plotted as a function of time for (a) $\Delta = 0.8J$ and (b) $\Delta = 1.0J$ with $L_S = 10$ and $L_B = 200$. We start from a filled state, and the bath is non-interacting. The early time data [$t$ up to $O(1/J)$], matches exactly with the analytical solutions given by Eq.~\eqref{eq:svn_early_time} for $\svn$ and Eqs.~\eqref{eq:Nbath_early_time} and \eqref{eq:var_Nbath_early_time} for $\expval{N_{\rm bath}}$ and $var(N_{\rm bath})$ respectively. In (c) and (d), we plot the density profiles for different time instances for $\Delta = 0.8J$ and $1.0J$, respectively.}
    \label{fig:non_int_dynamics}
\end{figure*}

\begin{figure*}[t]
    \centering
    \includegraphics[width=1.0\linewidth]{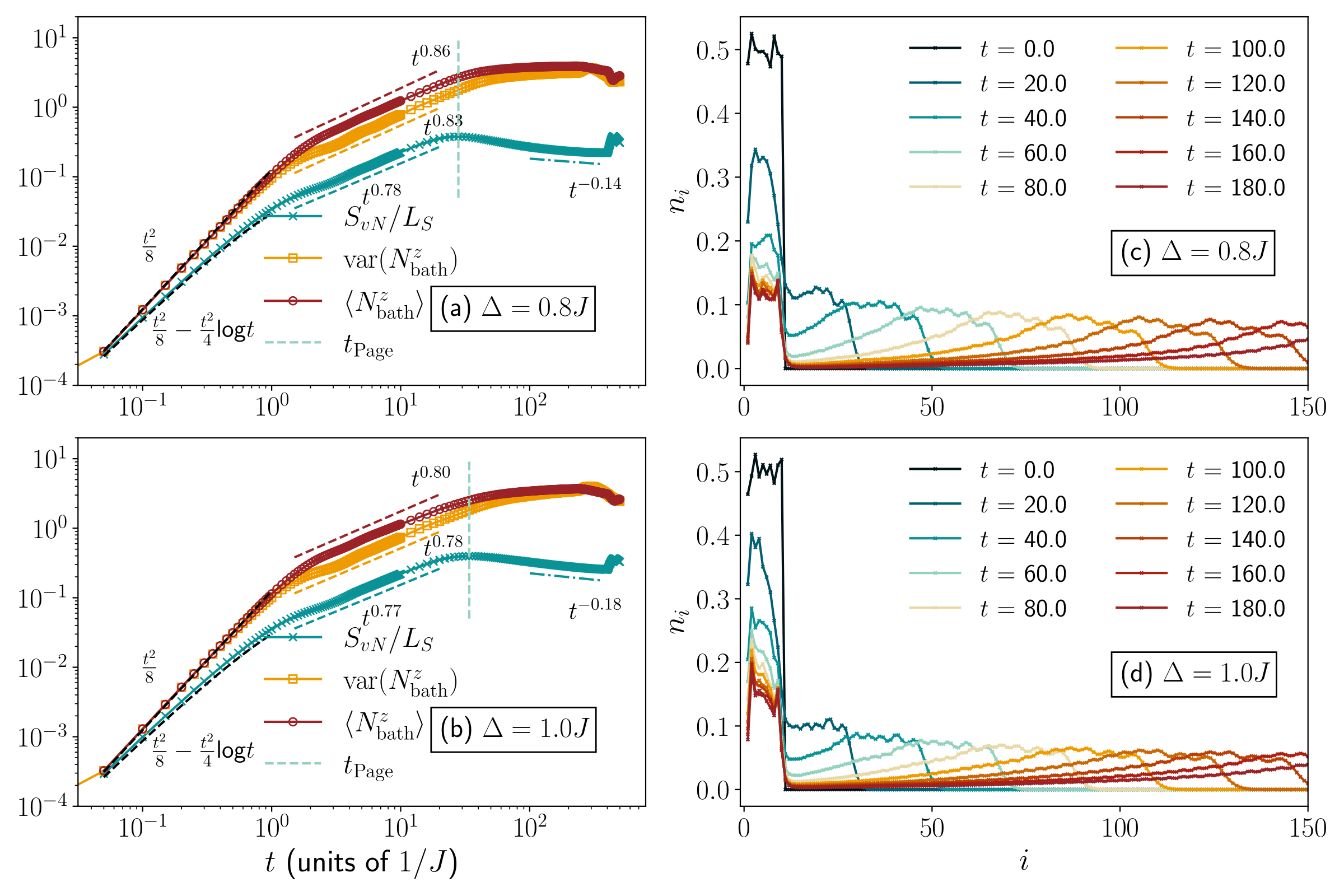}
    \caption{The $S_{vN}(t)$, the mean $\expval{N_{\rm bath}}$, and the fluctuation $var(N_{\rm bath})$ of the particle number of the bath, for an integrable interacting system are plotted as a function of time for (a) $\Delta = 0.8J$ and (b) $\Delta = 1.0J$ with $L_S = 10$ and $L_B = 200$. We start from a high Boltzmann entropy state, and the bath is non-interacting. The early time data [$t$ up to $O(1/J)$] for $\expval{N_{\rm bath}}$ and $var(N_{\rm bath})$, matches exactly with the analytical solutions given by Eqs.~\eqref{eq:inf_Nbath_early_time} and \eqref{eq:inf_var_Nbath_early_time} respectively. However, for $\svn$, we do not have an analytical expression. Nonetheless, the expression $t^2/8 - (t^2/4)\log t$ seems to be a good fit. In (c) and (d), we plot the density profiles for different time instances for $\Delta = 0.8J$ and $1.0J,$ respectively. }
    \label{fig:non_int_high_temp_dynamics}
\end{figure*}

In this appendix, we report the results when the system is connected to a non-interacting bath, that is, $\Delta = 0$ in Eqs.~\eqref{eq:int_bath_hamil} and \eqref{eq:int_sys_bath}. Recall that in the main text [Sec.~\ref{sec:numerical_results}], the bath was always interacting and integrable with $\Delta \neq 0$.

\textit{Filled (spin-polarized) initial state:} We start from a filled state [Eq.~\eqref{eq:sys_pol_init}]. At $t = 0$, the state of the full chain is given by $\ket{\psi(0)} = \ket{\psi_{\rm sys}(0)}\otimes \ket{\psi_{\rm bath}(0)}$, where $\ket{\psi_{\rm bath}(0)}$ is given by Eq.~\eqref{eq:micro_bath_pol_init}. In Fig.~\ref{fig:non_int_dynamics}, we plot the entanglement entropy $S_{vN}(t)$, the mean $\expval{N_{\rm bath}}$, and the fluctuation $var(N_{\rm bath})$ of the particle number of the bath, for the integrable case ($J' = 0$), as a function of time for (a) $\Delta = 0.8J$ and (b) $\Delta = 1.0J$ with $L_S = 10$ and $L_B = 200$. Similar to the case in Sec.~\ref{sec:integrable_numerics} in the main text, we find three prominent time regimes in the dynamics. In the first regime, up to $t\sim O(1/J)$, the numerics agree with the analytical solutions given in Eqs.~\eqref{eq:svn_early_time}, \eqref{eq:Nbath_early_time}, and \eqref{eq:var_Nbath_early_time} in the main text. Although these equations were derived for an integrable interacting bath, as we have earlier pointed out, in this time regime [$t$ up to $O(1/J)$], the dynamics are independent of the interaction. In the intermediate regime, for $\Delta = 0.8J$ the three quantities grow as 
\begin{align}
    \svn \sim t^{0.62},\, \expval{N_{\rm bath}} \sim t^{0.80}, \, var(N_{\rm bath}) \sim t^{0.78},
    \label{eq:non_int_int_regime_results1}
\end{align}
while for $\Delta = 1.0J$
\begin{align}
    \svn \sim t^{0.56},\, \expval{N_{\rm bath}} \sim t^{0.71}, \, var(N_{\rm bath}) \sim t^{0.75}.
    \label{eq:non_int_int_regime_results2}
\end{align}
Notice that the entanglement grows faster than the case when the system is connected to an interacting bath [see Eqs.~\eqref{eq:int_regime_results1} and \eqref{eq:int_regime_results2}]. At the Page time, $0.75$ and $0.65$ fractions of the particles have decayed into the bath for $\Delta = 0.8J$ and $\Delta = 1.0J$ respectively. After the Page time, for $\Delta = 0.8J$, the $\svn$ falls off as $1/t^{0.58}$ and $1/t^{0.5}$ for $\Delta = 1.0$. In Fig.~\ref{fig:non_int_dynamics} (c) and (d) we plot the density profiles at different time instances.

\textit{High entropy initial state:} Similar to the cases discussed in the main text [Sec.~\ref{sec:integrable_numerics}], we investigate the dynamics when the system is initialized in a high Boltzmann entropy state. In Fig.~\ref{fig:non_int_high_temp_dynamics}, we plot the entanglement entropy $S_{vN}(t)$, the mean $\expval{N_{\rm bath}}$, and the fluctuation $var(N_{\rm bath})$ of the particle number of the bath, for the integrable case ($J' = 0$), as a function of time for (a) $\Delta = 0.8J$ and (b) $\Delta = 1.0J$ with $L_S = 10$ and $L_B = 200$, starting from a high entropy state. In the first regime [up to $t \sim O(1/J)$], we find that the dynamics of these quantities have similar growth laws, but with different pre-factors,  compared to the case when the system is in the zero entropy filled state (Sec.~\ref{sec:early_time_integrable}). In particular, $\expval{N_{\rm bath}}$ and $var(N_{\rm bath})$ in this time regime are given by Eq.~\eqref{eq:inf_Nbath_early_time} and \eqref{eq:inf_var_Nbath_early_time} respectively. In the intermediate regime, for $\Delta = 0.8J$, the above three quantities grow as
\begin{align}
    \svn \sim t^{0.78},\, \expval{N_{\rm bath}} \sim t^{0.86}, \, var(N_{\rm bath}) \sim t^{0.83},
    \label{eq:heNN8_app}
\end{align}
while for $\Delta = 1.0J$
\begin{align}
    \svn \sim t^{0.77},\, \expval{N_{\rm bath}} \sim t^{0.80}, \, var(N_{\rm bath}) \sim t^{0.78}.
    \label{eq:heNN1_app}
\end{align}
In this case, like when the system is connected to an interacting bath (Sec.~\ref{sec:integrable_numerics}), the entanglement grows faster in the intermediate regime up to the Page time for the high entropy initial state, compared to the zero entropy initial state. Notice that the exponents are not much different from the case when the system is connected to an interacting bath and initiated in a high entropy state. At the Page time, $0.54$ and $0.50$ fractions of the particles have decayed into the bath for $\Delta = 0.8J$ and $\Delta = 1.0J$ respectively. After the Page time, for $\Delta = 0.8J$, the $\svn$ falls off as $1/t^{0.14}$ and $1/t^{0.18}$ for $\Delta = 1.0$. In Fig.~\ref{fig:non_int_high_temp_dynamics} (c) and (d) we plot the density profiles at different time instances. In a recent study \cite{RJFeb2025}, the authors have studied the dynamics of entanglement when an interacting Fermionic system is connected to a non-interacting Fermionic bath.

\begin{figure*}
    \centering
    \includegraphics[width=1.0\linewidth]{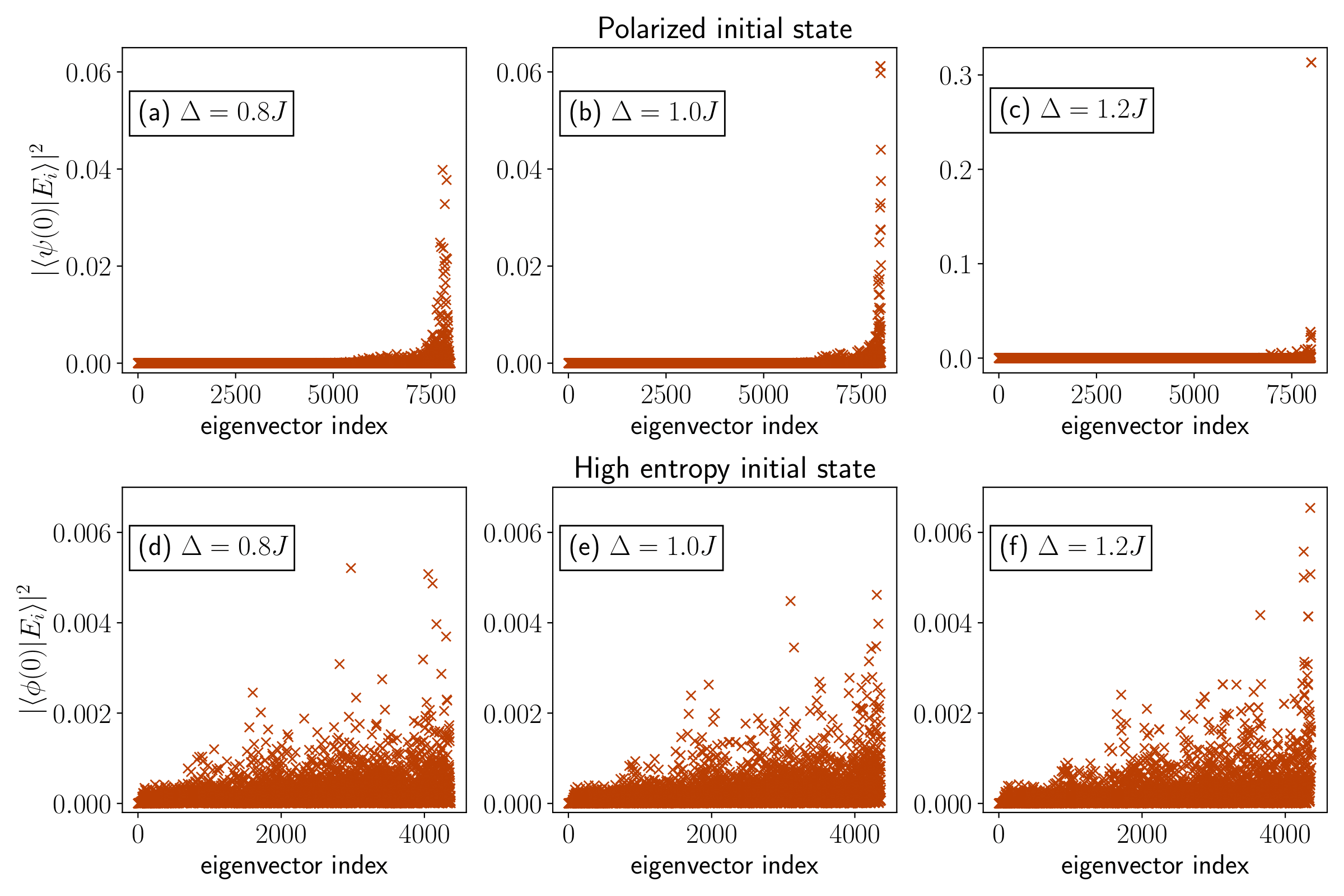}
    \caption{The overlap of the initial state with all the eigenstates ($\{E_i\}$) of the system and the bath is plotted for (a,d) $\Delta = 0.8J$, (b,e) $\Delta = 1.0J$ and (c,f) $\Delta = 1.2J$. In (a-c) we consider the polarized initial state $\ket{\psi(0)} = \ket{\psi_{\rm sys}(0)}\otimes \ket{\psi_{\rm bath}(0)}$ [Eqs.~\eqref{eq:sys_pol_init} and \eqref{eq:micro_bath_pol_init}] and in (d-f) we consider the high entropy initial state $\ket{\phi(0)}$ [Eq.~\eqref{eq:inf_temp_state}]. The system is integrable ($J' = 0$) with $L_S = 10$, and $L_B = 6$.}
    \label{fig:overlap}
\end{figure*}

\begin{figure*}
    \centering
    \includegraphics[width=1.0\linewidth]{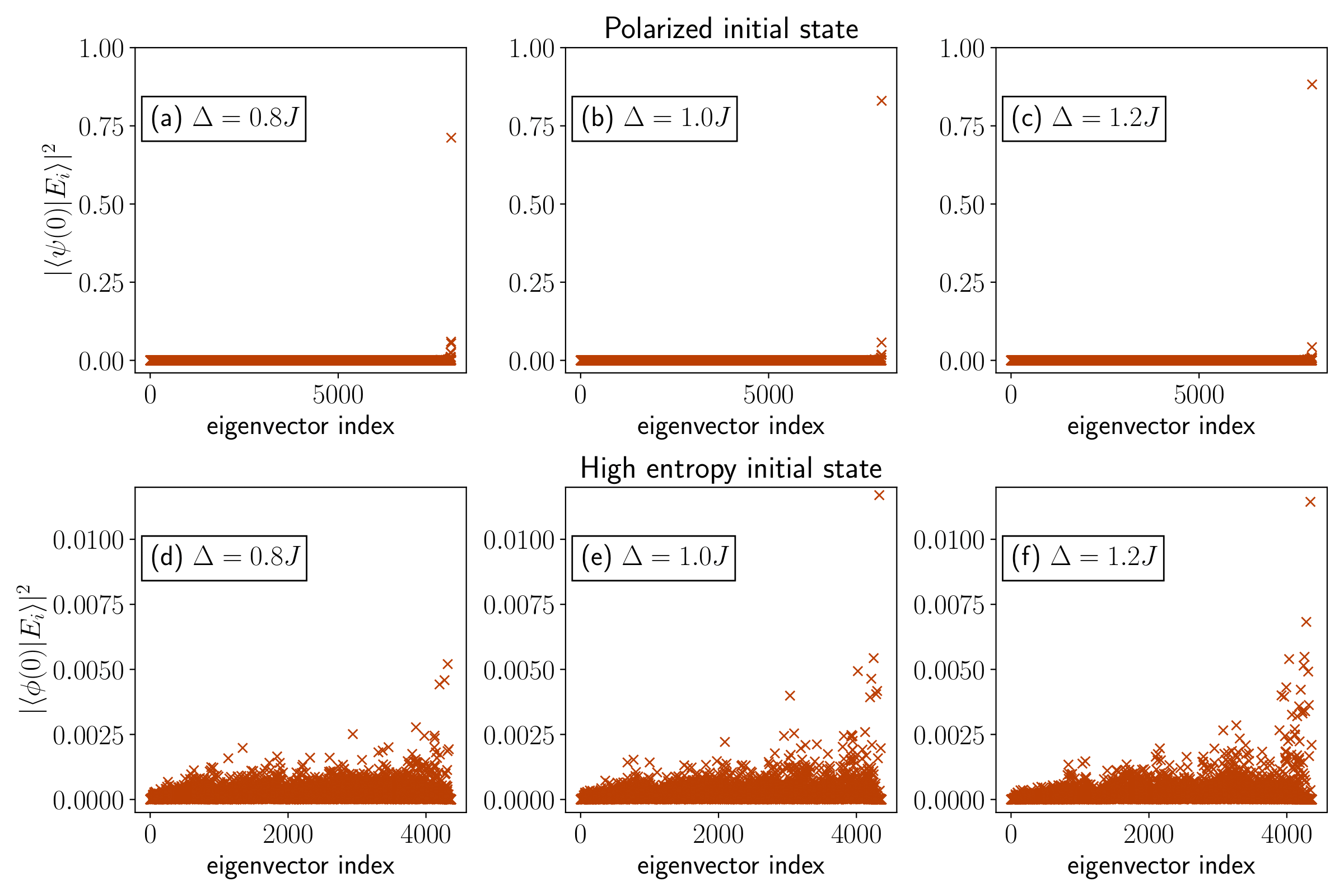}
    \caption{The overlap of the initial state with all the eigenstates ($\{E_i\}$) of the system and the bath is plotted for (a,d) $\Delta = 0.8J$, (b,e) $\Delta = 1.0J$ and (c,f) $\Delta = 1.2J$. In (a-c) we consider the polarized initial state $\ket{\psi(0)} = \ket{\psi_{\rm sys}(0)}\otimes \ket{\psi_{\rm bath}(0)}$ [Eqs.~\eqref{eq:sys_pol_init} and \eqref{eq:micro_bath_pol_init}] and in (d-f) we consider the high entropy initial state $\ket{\phi(0)}$ [Eq.~\eqref{eq:inf_temp_state}]. The system is non-integrable ($J' = 1$) with $L_S = 10$, and $L_B = 6$.}
    \label{fig:overlap_NNN}
\end{figure*}

\section{Initial state overlap with the spectrum}
\label{app:overlap}
In this appendix, we discuss the overlap of the initial states considered in the study with the eigen spectrum of the models considered in the main text. Given the large size of the bath ($L_B = 200$), it is unfeasible to compute the entire spectrum of the system and the bath. However, since we want to compute the overlap of the initial state, where the particles are localized in the system, it is sufficient to consider only a few sites in the bath. We set $L_S = 10$ and $L_B = 6$. First, we consider the case when the system is integrable ($J' = 0$). In Fig.~\ref{fig:overlap} (a-c), we plot the overlap $\abs{\bra{\psi(0)}\ket{E_i}}^2$ of the polarized initial state $\ket{\psi(0)} = \ket{\psi_{\rm sys}(0)}\otimes \ket{\psi_{\rm bath}(0)}$ [Eqs.~\eqref{eq:sys_pol_init} and \eqref{eq:micro_bath_pol_init}] with the eigenvectors $\{\ket{E_i}\}$ of the Hamiltonian of the system and bath, when $J'=0$, for $\Delta = 0.8J$, $\Delta = 1.0J$, and $\Delta = 1.2J$, respectively. Note that the polarized initial state has a high overlap with one of the eigenstates for $\Delta = 1.2J$, which leads to the freezing of the dynamics. On the other hand, for $\Delta = 0.8J$ and $\Delta = 1.0J$, the initial state has support over a band of eigenstates. In Fig.~\ref{fig:overlap} (d-f), we plot the overlap $\abs{\bra{\phi(0)}\ket{E_i}}^2$ of the high entropy initial state $\ket{\phi(0)}$ [Eq.~\eqref{eq:inf_temp_state}], with the eigenvectors for $\Delta = 0.8J$, $\Delta = 1.0J$, and $\Delta = 1.2J$, respectively. In this case, we notice that the state has support over a band of the eigenstates for any $\Delta$.

Next, we consider the case when the system is non-integrable ($J' = 1$). In Fig.~\ref{fig:overlap_NNN} (a-c), we plot the overlap $\abs{\bra{\psi(0)}\ket{E_i}}^2$ of the polarized initial state $\ket{\psi(0)} = \ket{\psi_{\rm sys}(0)}\otimes \ket{\psi_{\rm bath}(0)}$ [Eqs.~\eqref{eq:sys_pol_init} and \eqref{eq:micro_bath_pol_init}] with the eigenvectors $\{\ket{E_i}\}$ of the Hamiltonian of the system and bath, when $J'=1$, for $\Delta = 0.8J$, $\Delta = 1.0J$, and $\Delta = 1.2J$, respectively. In this case, for all $\Delta$, we find that the polarized initial state is very close to one of the eigenstates, which leads to the freezing of the dynamics as reported in Fig.~\ref{fig:NNN_dynamics}. In Fig.~\ref{fig:overlap_NNN} (d-f), we plot the overlap $\abs{\bra{\phi(0)}\ket{E_i}}^2$ of the high entropy initial state $\ket{\phi(0)}$ [Eq.~\eqref{eq:inf_temp_state}], with the eigenvectors for $\Delta = 0.8J$, $\Delta = 1.0J$, and $\Delta = 1.2J$, respectively. In this case, we notice that the state has support over a band of the eigenstates for any $\Delta$.

\section{Numerical details of TEBD}
\label{app:TEBD}
In this appendix, we discuss the time evolution block decimation (TEBD) algorithm. To study the dynamical properties of the one-dimensional closed chain whose Hamiltonian is of the form
\begin{equation}
H = \sum_{n = 1}^{L-1}h_{n,n+1},
\label{eq:1D_NN_chain}
\end{equation}
where $h_{n,n+1}$ is a two-site operator acting on sites $n$ and $n+1$, we use the TEBD algorithm \cite{USJan2011}. We represent the state of such a system as a matrix product state (MPS) \cite{USJan2011}
\begin{equation}
    \ket{\psi} = \sum_{j_1,j_2,\dots,j_{L}=1}^DB^{(1)j_1}B^{(2)j_2}\dots B^{(L)j_{L}}\ket{j_1,j_2,\dots,j_{L}},
    \label{eq:MPS}
\end{equation}
where $B^{(n)j_n}$ are matrices at site $n$ and $j_n$'s are the physical indices of the system, each having a local Hilbert space dimension $D$. The dimensions of these matrices, also known as the bond dimension, grow exponentially with system size, as one approaches the center of the chain. We set a cutoff $\chi$ in the bond dimension of these matrices, which leads to a loss of information in the state $\ket{\psi}$. To ensure that the error due to this loss is not significant, we check for the convergence of the dynamics of observables with increasing $\chi$. To evolve the system, we implement a fourth-order Trotter decomposition of the evolution operator $\exp(-\ii \delta t H)$ where $H$ is given by Eq.~\eqref{eq:1D_NN_chain} as
\begin{equation}
    e^{-\ii \delta t H} = \U(\delta t_1)\U(\delta t_1)\U(\delta t_2)\U(\delta t_1)\U(\delta t_1) + {\cal O}(\delta t^5),
    \label{eq:order5_trotter}
\end{equation}
where
\begin{equation}
    \U(\delta t_i) = e^{-\ii H_{\rm odd}\delta t_i/2}e^{-\ii H_{\rm even}\delta t_i}e^{-\ii H_{\rm odd}\delta t_i/2},
    \label{eq:order2_trotter}
\end{equation}
and
\begin{eqnarray}
    \delta t_1 &=& \frac{\delta t}{4-4^{1/3}},\\
    \delta t_2 &=& \delta t - 4\delta t_1.
\end{eqnarray}
In Eq.~\eqref{eq:order2_trotter}, $H_{\rm odd} = \sum_{n\in \text{odd}} h_{n,n+1}$ and $H_{\rm even} = \sum_{n\in \text{even}} h_{n,n+1}$. The state at time $t = m \delta t$ is given by
\begin{equation}
    \ket{\psi(t)} \approx [\U(\delta t_1)\U(\delta t_1)\U(\delta t_2)\U(\delta t_1)\U(\delta t_1)]^m\ket{\psi(0)}.
\end{equation}

For the case where the system is attached to a bath, we consider the entire chain of the system and the bath as a closed system. When $J' = 0$, that is, when the system is integrable, and the bath is interacting (and integrable), the Hamiltonian in Eq.~\eqref{eq:sys_micro_bath_hamil} is already of the form given in Eq.~\eqref{eq:1D_NN_chain} with
\begin{eqnarray}
    h_{n,n+1} = J\left(S^x_{n}S^x_{n+1} + S^y_{n}S^y_{n+1} + \Delta S^z_{n}S^z_{n+1}\right),
    \label{eq:two_site_xxz}
\end{eqnarray}
$L = L_S + L_B$, and $D = 2$. When the same system is connected to a non-interacting bath, the two-site energy operator is given by
\begin{align}
    h_{n,n+1} \!= 
    \begin{cases}
        J\big(S^x_{n}S^x_{n+1} + S^y_{n}S^y_{n+1} +\Delta S^z_{n}S^z_{n+1}\big),&n\in [L_S-1,0] \\
        J\left(S^x_{n}S^x_{n+1} + S^y_{n}S^y_{n+1}\right), & n\in[1,L_B-1]
    \end{cases}.
\end{align}

\begin{figure}[t]
    \centering
    \includegraphics[width=1.0\linewidth]{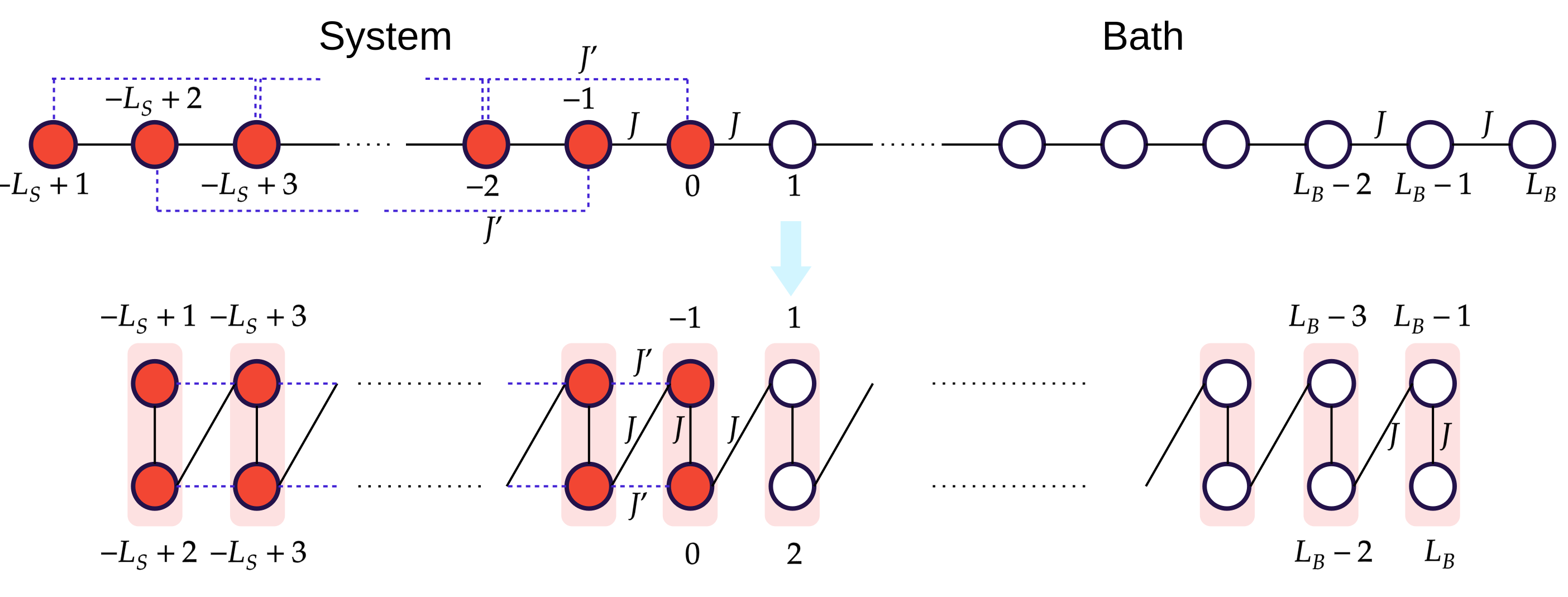}
    \caption{A schematic representation of how the one-dimensional chain with NNN coupling is redrawn as a ladder with NN coupling. Each local site on the ladder site is highlighted with a pink box.}
    \label{fig:ladder_lattice}
\end{figure}

When $J' \neq 0$, the Hamiltonian has NNN coupling. To transform this Hamiltonian into the form in Eq.~\eqref{eq:1D_NN_chain}, we redraw the lattice as a ladder [see Fig.~\ref{fig:ladder_lattice}], where two neighboring sites in the linear lattice combine to form a single site in the ladder lattice of length $(L_S + L_B)/2$ (we assume both $L_S$ and $L_B$ to be even for simplicity) with $D = 4$. We label the spin operators on top with subscript $u$ and the one at the bottom with subscript $d$. As evident from Fig.~\ref{fig:ladder_lattice}, such a system has only NN coupling, and the two-site energy operator is given by
\begin{widetext}
    \begin{align}
    h_{n,n+1} = 
    \begin{cases}
        J[(S^x_uS^x_d)_n + (S^y_uS^y_d)_n + \Delta (S^z_uS^z_d)_n]\otimes \I_{n+1} \\
         + J[(\I_uS^x_d)_n\otimes(S^x_u\I_d)_{n+1} + (\I_uS^y_d)_n\otimes(S^y_u\I_d)_{n+1} + \Delta(\I_uS^z_d)_n\otimes(S^z_u\I_d)_{n+1}], & n \in \left[0,\frac{L_B}{2}-1\right] \\ \\
        J[(S^x_uS^x_d)_n + (S^y_uS^y_d)_n + \Delta (S^z_uS^z_d)_n]\otimes \I_{n+1} \\
         + J[(\I_uS^x_d)_n\otimes(S^x_u\I_d)_{n+1} + (\I_uS^y_d)_n\otimes(S^y_u\I_d)_{n+1} + \Delta(\I_uS^z_d)_n\otimes(S^z_u\I_d)_{n+1}]\\
         + J'[(S^x_u\I_d)_n\otimes(S^x_u\I_d)_{n+1}) + (S^y_u\I_d)_n\otimes(S^y_u\I_d)_{n+1}) + \Delta (S^z_u\I_d)_n\otimes(S^z_u\I_d)_{n+1})]\\
         + J'[(\I_uS^x_d)_n\otimes(I_uS^x_d)_{n+1} + (\I_uS^y_d)_n\otimes(\I_uS^y_d)_{n+1} + \Delta (\I_uS^z_d)_n\otimes(\I_uS^z_d)_{n+1}], & n \in \left[-\left(\frac{L_S}{2}-1\right),-1\right]
    \end{cases}.
\end{align}
\end{widetext}

\begin{figure}[t]
    \centering
    \includegraphics[width=1.\linewidth]{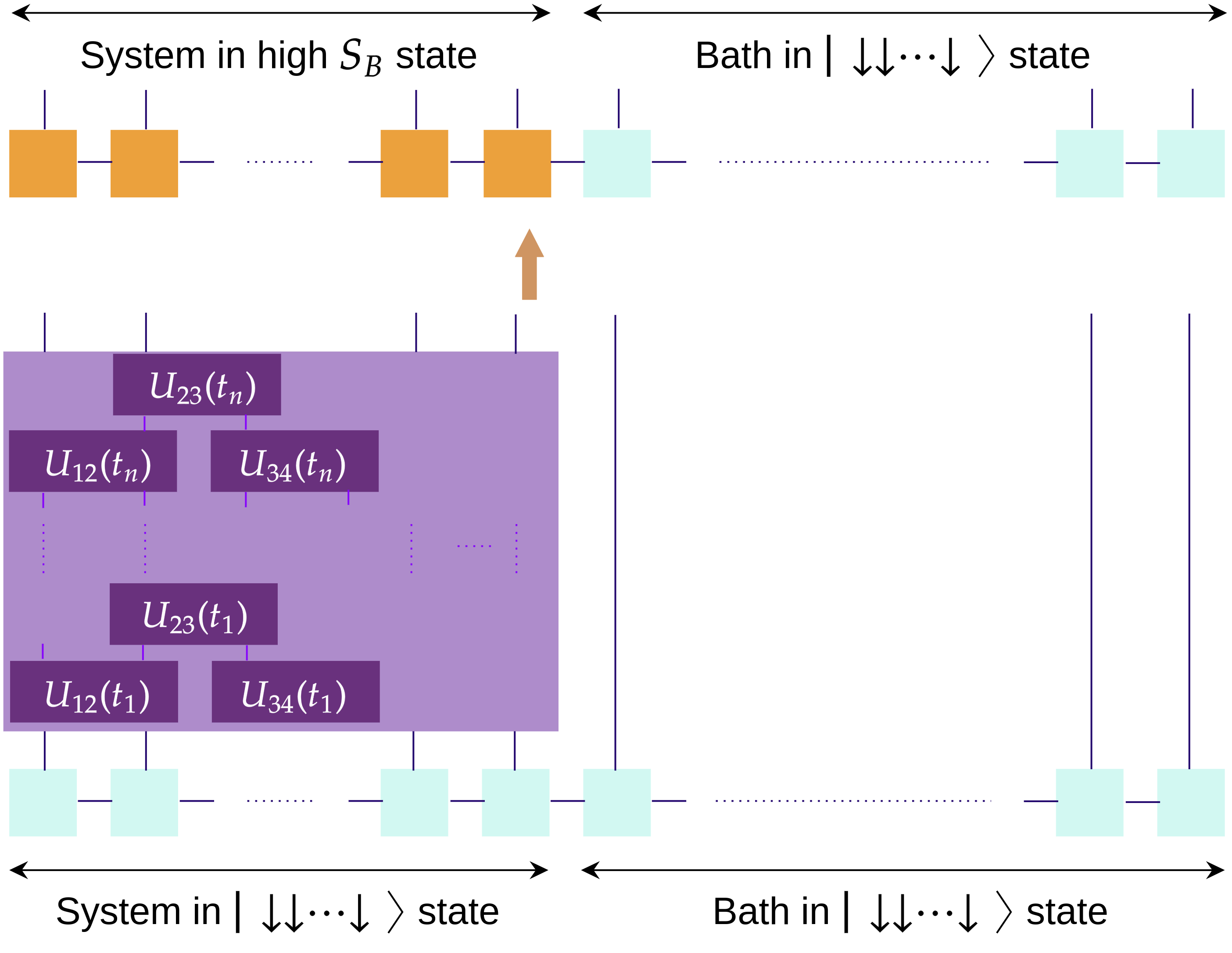}
    \caption{Schematic representation of preparing a state with high Boltzmann entropy in the MPS representation. The Haar random 2-site gates $U_{j,j+1}(t_k)$ acts on the system sites $j$ and $j+1$ at step $k$. The depth of the circuit $n \sim L_S$. The bath sites are left untouched.}
    \label{fig:haar}
\end{figure}

To prepare a state with high Boltzmann entropy in the MPS form, we apply a series of Haar random two-site unitary gates to the MPS of a simple product state. These quantum two-site gates are randomly chosen from Haar measure of the unitary group. This is done such that it is ensured that each unitary gate is selected with equal probability. The depth of the circuit is taken to be $\mathcal{O}(L_S)$. The Haar random two-site gates are only applied on the system sites, whereas the bath sites, in the empty state, are left untouched. This transformation simulates a random quantum circuit localized around the system sites and leads to a state where the system has high Boltzmann entropy and the bath remains in the zero Boltzmann entropy state [Eq.~\eqref{eq:micro_bath_pol_init}]. A schematic representation of the algorithm is given in Fig.~\ref{fig:haar}.

\bibliography{bibliography}
\end{document}